\documentclass{article}

\usepackage{arxiv}

\usepackage[utf8]{inputenc} 
\usepackage[T1]{fontenc}    
\usepackage{hyperref}       
\usepackage{url}            
\usepackage{booktabs}       
\usepackage{nicefrac}       
\usepackage{microtype}      
\usepackage{lipsum}		
\usepackage{natbib}
\usepackage{doi}

\usepackage{cite}
\usepackage{amsmath,amssymb,amsfonts}
\usepackage{bm} 
\usepackage{graphicx}
\usepackage{caption}
\usepackage{multirow}

\usepackage{tabularx} 

\usepackage{subfigure}

\usepackage{cleveref}

\usepackage{makecell}

\usepackage{algorithm,algpseudocode}
\algnewcommand{\Inputs}[1]{%
  \State \textbf{Inputs:}
  \Statex \hspace*{\algorithmicindent}\parbox[t]{.8\linewidth}{\raggedright #1}
}
\algnewcommand{\Initialize}[1]{%
  \State \textbf{Initialize:}
  \Statex \hspace*{\algorithmicindent}\parbox[t]{.8\linewidth}{\raggedright #1}
}

\usepackage{xcolor}

\title{A Parallel Tempering Approach for Efficient Exploration of the Verification Tradespace in Engineered Systems}

\author{ {Peng~Xu} \\
	Grado Department of Industrial and Systems Engineering\\
	Virginia Tech\\
	Blacskburg, VA 24060 \\
	\texttt{xupeng@vt.edu} \\
	\And
	{Alejandro~Salado} \\
	Department of Systems and Industrial Engineering\\
	The University of Arizona\\
	Tucson, AZ 85721 \\
	\texttt{alejandrosalado@arizona.edu} \\
	\And
	{Xinwei~Deng} \\
	Department of Statistics\\
	Virginia Tech\\
	Blacskburg, VA 24060 \\
	\texttt{xdeng@vt.edu} \\
}

\renewcommand{\shorttitle}{\textit{arXiv} Template}

\hypersetup{
pdftitle={A Parallel Tempering Approach for Efficient Exploration of the Verification Tradespace in Engineered Systems},
pdfsubject={EESS},
pdfauthor={Peng~Xu, Alejandro~Salado, and~Xinwei~Deng},
pdfkeywords={verification strategy, engineered system, set-based design, tradespace exploration, tree search, bayesian network, parallel tempering},
}

\begin{document}
\maketitle

\begin{abstract}
	Verification is a critical process in the development of engineered systems. Through verification, engineers gain confidence in the correct functionality of the system before it is deployed into operation. Traditionally, verification strategies are fixed at the beginning of the system's development and verification activities are executed as the development progresses. Such an approach appears to give inferior results as the selection of the verification activities does not leverage information gained through the system's development process. In contrast, a set-based design approach to verification, where verification activities are dynamically selected as the system's development progresses, has been shown to provide superior results. However, its application under realistic engineering scenarios remains unproven due to the large size of the verification tradespace. In this work, we propose a parallel tempering approach (PTA) to efficiently explore the verification tradespace. First, we formulate exploration of the verification tradespace as a tree search problem. Second, we design a parallel tempering (PT) algorithm by simulating several replicas of the verification process at different temperatures to obtain a near-optimal result. Third, We apply the PT algorithm to all possible verification states to dynamically identify near-optimal results. The effectiveness of the proposed PTA is evaluated on a partial model of a notional satellite optical instrument.
\end{abstract}

\keywords{verification strategy \and engineered system \and set-based design \and tradespace exploration \and tree search \and  Bayesian network \and parallel tempering}

\begin{center}
  \large\bfseries  
  This work has been submitted to the IEEE for possible publication. Copyright may be transferred without notice, after which this version may no longer be accessible.
\end{center}

\newpage

\section{Introduction}
\label{sec:introduction}
System verification, defined as the process of evaluating whether a system or its components fulfill their requirements, is generally executed during the development of engineered systems~\citep{Engel2010,grobelna2016design,machin2016smof}. System verification is often planned and implemented as a strategy of the verification activities, which can be executed at different developmental phases and on different system configurations~\citep{Salado2018}. A well-designed verification strategy can contribute to the expected utility of a system in multiple ways such as by shaping beliefs about a system exhibiting certain characteristics, consuming resources, or informing the need for certain design features~\citep{Salado2018properties}.

In current practice, verification strategies are usually defined and fixed at the beginning of system development by allocating and committing the resources necessary to execute the planned verification activities (VA) throughout the development process~\citep{Engel2010}. Several optimization algorithms have been proposed to support this approach (e.g., ~\citep{Engel2010,engel2003methodology,barad2006optimizing,lv2014adaptive,xiao2017adaptive}), all of which rely on the assumption that the value (or information) provided by an individual VA is a constant. However, the information generated by a verification activity and, hence, its value and the necessity to perform it, is a function of the results of VAs that have previously been performed ~\citep{Salado2018}. In other words, as the system development progresses, VAs that were initially considered necessary may become unnecessary given intermediate verification results and vice versa~\citep{xu2019concept}. Therefore, defining and fixing a verification strategy early in system development yields suboptimal value~\citep{xu2019concept}. Instead, dynamically defining verification strategies, where the selection and execution of verification activities change as verification results from previous verification activities are obtained, yield higher value~\citep{xu2019concept}. In essence, with dynamic verification strategies, a VA is only performed if worthy, not because it was originally committed at the beginning of the system development. 

The set-based design (SBD)~\citep{ward1995second,ward1995toyota} is a promising technique to define dynamic verification strategies~\citep{xu2019concept}. 
However, operationalization concerns remain in the implementation of such a technique. 
In particular, as the effect of each activity is influenced by both the results of previous activities and the choice of future activities, verification strategies cannot be decomposed into basic activities that are assessed independently. Instead, whole verification strategies should be considered when assessing how valuable each VA is. However, as the number of possible verification activities increases, the magnitude of the resulting possible verification strategies (i.e., the verification tradespace) becomes so large that using enumeration to identify the best verification strategy becomes infeasible due to the curse of dimensionality~\citep{ernest2003dynamic}. Therefore, operationalizing the design and use of dynamic verification strategies requires development of feasible exploration approaches for large verification tradespaces.

To overcome these problems, this paper presents a feasible exploration framework based on a parallel tempering approach (PTA) that enables the application of SBD to dynamically define verification strategies in large verification tradespaces. First, we reformulate the verification tradespace as a tree space rather than as a path space. In this way, the exploration approach becomes a process that identifies the near-optimal VA at each state to narrow down the set of verification strategies. Next, we design a parallel tempering (PT) algorithm that runs a series of system replicas to find the near-optimal foresight verification tree (FVT), where the root node of the FVT is set as the near-optimal VA at each state. Finally, the exploration results for all states are collected as a hindsight verification tree (HVT) to evaluate the performance of our proposed method.

The remainder of this paper is organized as follows. 
Section~\ref{sec:background} reviews the background about SBD and PT algorithm. Section~\ref{sec:methodology} describes the proposed methodology to make the exploration of the verification tradespace feasible. Section~\ref{sec:experiment} presents our experiments and related analyses, and, finally, a summary of the conclusions of this paper is presented in Section~\ref{sec:Conclusion}.

\section{Background}
\label{sec:background}

\subsection{Set-based Design}
\label{subsec:setbaseddesign}

There is often a lack of knowledge about a system at the beginning of development of that system~\citep{blanchard1990systems}. Such a lack of knowledge motivates the emergence of SBD~\citep{ward1995second,ward1995toyota}. SBD is built on the principle of working simultaneously with a plethora of design alternatives instead of converging quickly to a single option. As knowledge about the system increases during system development, suboptimal alternatives are discarded until a preferred one remains~\citep{bernstein1998design}. SBD has been successfully applied in multiple applications, including structural design~\citep{miller2018design}, naval systems~\citep{singer2009set}, multiplate clutch systems~\citep{canbaz2014resolving}, and 3D metal forming processes~\citep{schjott2019using}, among others. SBD has also been shown to strengthen the performance of tradespace exploration~\citep{small2018uav}. In particular, whereas tradespace exploration can identify numerous solutions in the initial design set~\citep{Ross200511}, SBD can reduce the burden of finding the optimal choice in early stages. For example, Specking et al.~\citep{specking2018early} proposed an integrated framework for an unmanned aerial vehicle case and showed how SBD was able to find a larger set of feasible designs early in the design process compared to traditional methods.

In the field of system verification, Xu and Salado~\citep{xu2019concept} proposed using SBD to enable the design of dynamic verification strategies as verification results become available. In essence, an engineering team would work with a set of verification paths instead of only a single verification path. The set is formed by those verification paths that are optimal for the different results that future verification activities might yield. Once a VA is executed and its results are known, the values of the verification paths in the set are updated, which makes some of them suboptimal. These suboptimal paths are then removed from the set, which continuously shrinks as system development progresses. In Xu and Salado's concept paper, identification of optimal paths within the verification tradespace was performed using enumeration. Hence, the computational approach is not scalable and cannot address the design of verification strategies for more realistic systems. This is the main shortcoming that is addressed in this paper.

\subsection{Bayesian Networks}
\label{subsec:bayesnet}

A Bayesian network (BN) is a probabilistic graphical model that represents a set of random variables and their conditional dependencies via a directed acyclic graph~\citep{Cai2017}. BNs have been used as fundamental tools for verifying engineered systems~\citep{Salado2018}. In particular, system parameters to be verified and verification activities to be performed are modeled as nodes in the BN, where edges represent their information influence. Then, the information dependency of verification activities is captured by joint distribution of the BN.

The execution of a verification strategy is, hence, modeled as a Bayesian inference process~\citep{koller2009probabilistic} that captures the way in which engineers build confidence in the state of the system as verification evidence becomes available~\citep{Salado2019}. Realization of this Bayesian inference process consists of three steps: (1) A network structure is built that captures the causal relationships between all system parameters and verification activities; (2) The network nodes are assigned with prior distributions generated through knowledge elicitation; and (3) Activity results are collected during the verification process, enabling updates of posterior distributions of the state of the system. 

\subsection{Parallel Tempering}
\label{subsec:paralleltempering}
For problems where finding an optimal solution is very difficult or not practical, heuristic methods are often used to facilitate the process of finding a satisfactory solution~\citep{wikiHeuristic,zhang2013optimal,xiong1998scheduling}. Parallel Tempering (PT) is a heuristic method originally devised by Swendsen and Wang~\citep{swendsen1986replica}. This method simulates $M$ replicates $\{\Omega(\Psi_m)\}$ of the original system of interest simultaneously. 
Each replica is assigned with a different temperature $\Psi_m$ that originally represents physical temperature in molecular dynamics~\citep{swendsen1986replica}. For ordinary systems, temperatures are used as hyperparameters of the PT algorithm that have a direct impact on the acceptance probability of the Monte Carlo process. If the temperature of a replica is high, the replica can accept new samples in a larger solution space. Even though PT has $M$ replicas, which requires $M$ times more computational effort than a standard, single-temperature replica simulation, PT is over $1/M$ times more efficient than the latter because it allows the lower temperature system to jump out of its regular region of sampling~\citep{Earl2005}. In addition, PT can make efficient use of large CPU clusters, since replicas can be simulated in parallel. Due to these benefits, this paper leverages the PT method to identify optimal verification strategies with limited computational resources.

The standard PT method consists of a two-level sampling process, a basic level and an advanced level, as shown in Algorithm~\ref{alg:ClassicPT}. Suppose there are M replicas $\{\Omega(\Psi_m)\}$ with their own configurations (i.e., $\Omega(\Psi_m) = x^i_m)$). In the basic level, a Markov chain Monte Carlo (MCMC) simulation~\citep{metropolis1953equation} would be run for each replica. In the advanced level, all pairs of two neighboring replicas may exchange their configurations $\{x^i_m,x^i_{m+1}\}$ with acceptance probability $p=min(1,exp(\Delta \beta \Delta E))$, where $\Delta \beta = \frac{1}{\Psi_m}-\frac{1}{\Psi_{m+1}}$, $\Delta E=E_m - E_{m+1}$, and $E_m$ is the performance metric of the configuration $x^i_m$. The probability is chosen in such a way that the exchange of replicas is reversible by satisfying the detailed balance condition~\citep{Wang2009}; hence, this condition is satisfied for the complete PT method. This is an inherent advantage of PT, since it can provide a desirable equilibrium state for the sampling process.

\begin{algorithm}
  \caption{Standard PT Algorithm}
  \label{alg:ClassicPT} 
  \begin{algorithmic}[1]
    \Inputs{$N_s, N_{it}, \{\Psi_m\}, m=1,\ldots,M$.}
    \Initialize{$\{\Omega(\Psi_m)\}:\Omega(\Psi_m) = x^i_m$.}
    \For{i = 1 to $N_s$}
      \For{m = 1 to M}
        \State \parbox[t]{200pt}{Apply the MCMC method to $\Omega(\Psi_m)$ for $N_{it}$ iterations.\strut}
      \EndFor
      \For{m = 1 to M-1}
        \State \parbox[t]{200pt}{Swap $x^i_m$ with $x^i_{m+1}$ with the probability $p=min(1,exp(\Delta\beta\Delta E))$.\strut}
      \EndFor
    \EndFor
  \end{algorithmic}
\end{algorithm}

Design of a PT method involves three major hyperparameters~\citep{Wang2009}: the total number of swaps $N_s$, the number of MCMC iterations $N_{it}$, and the set of temperatures $\{\Psi_m\}$. Several rules of thumb have been proposed for designing these parameters. First, $N_{it}$ should be large enough so that a replica reaches equilibrium after $(M-1)N_{it}$ steps~\citep{Wang2009}. In terms of temperatures, the highest temperature should be high enough that its corresponding replica can cross the whole solution space; the lowest temperature should be low enough that its replica approaches the local minimum, $min(E_m)$~\citep{Wang2009}. Moreover, an optimal allocation of temperatures should lead to a uniform acceptance probability for all pairs of neighboring replicas~\citep{rathore2005optimal}. A simplistic approach is provided in~\citep{kofke2002acceptance}, which states that a geometric progression of temperatures (i.e., $\frac{\Psi_{m+1}}{\Psi_m}=C$) for ordinary systems could result in uniform acceptance probabilities. The spacing between the temperatures should also be small enough that a sufficiently large acceptance probability is reached~\citep{Wang2009, kofke2002acceptance}. The optimal acceptance probability is recommended to be $20\%$ for the cases in ~\citep{rathore2005optimal}.

\section{Proposed Methodology to Design Verification Strategies}
\label{sec:methodology}

\subsection{Basic Model of a Verification Construct}
\label{subsec:verification model}
\begin{figure}[htbp]
\vspace{-5mm}
\centerline{\includegraphics[width=5cm]{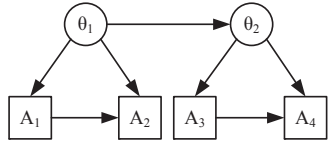}}
\caption{Illustration of an exemplar network}
\label{fig:examplenetwork}
\vspace{-5mm}
\end{figure}

In this work, the basic verification construct is modeled as a BN, which was also used in the previous study~\citep{Salado2018}, under the assumption that a BN can capture all confidence relationship between a complex system and its verification activities.
To illustrate the idea, an exemplar network is presented in Fig.~\ref{fig:examplenetwork}. 
System parameters are denoted by $\theta_k$ and verification activities by $A_i$. Then the confidence of the target parameter $P(\theta_1)$ can be deduced using the Bayesian inference.
For this paper, all nodes are considered binary: a system presents either error or no error, 
and a VA can yield either positive or negative results. 
Arrows represent information dependencies.

The verification process constitutes the execution of a verification strategy, i.e., a set of verification activities within $T$ time intervals that form a verification path. Each activity is conducted at the beginning of its corresponding time interval, i.e., $t = 0,...,T-1$ and can be conducted at most once in the verification process. Once the verification process is completed, the system will be deployed at the end $t=T$.  

\subsection{Valuation of Dynamic Verification Strategies}
\label{subsec:performancemeasurement}
In this study, the value of a verification path, $U$, is given by the summation of three factors. 
The first factor is the verification activity execution cost, $C_{A_i}$, which is a fixed amount of financial resources necessary to conduct a VA, $A_i$. The second factor is the rework cost, $C_{R_i}$, which represents the financial resources necessary to adjust the system when necessary. This cost is only incurred if rework activity, $R_i$, is triggered. 
The third factor is the system revenue, $B_k$, which is obtained once the system is deployed and operates correctly. 

While $C_{A_i}$ is linked to the execution of a VA, $C_{R_i}$ and $B_k$ depend on the evolution in confidence that the system is operating correctly as verification activities are performed. For simplicity, decision rules for the execution of rework activities and deployment of the system have been defined in this paper, regardless of their actual optimality within an expected utility framework~\citep{von1953theory}. In particular, a rework activity may be initiated when a VA, $A_i$, fails and the resulting confidence in the correct operation of the system falls below a predefined threshold $H_l$. $H_l$ mean the confidence threshold of rework activities. They are specific in the experiments and provided beforehand. Moreover, it is assumed that a rework activity raises confidence in the correct operation of the system to the level that would have been attained if the last verification activity before the rework had been successful. Similarly, we consider the system to be deployed when the confidence levels, $P(\theta_k)$, of the target parameters, $\theta_k$, reach or surpass certain thresholds. For simplicity and practicality, these rework and system deployment rules against confidence thresholds are predefined. 

Under these conditions, the expected value of a verification path at the end, $t = T$, is given by: 
\begin{equation} \label{eq:1}
\begin{aligned}
U(\mathbf{S}_{T}) = &\sum_k B_k P(\theta_k|\mathbf{S}_{T})\delta (P(\theta_k|\mathbf{S}_{T}) > H_u)\\
&-\sum_i C_{A_i}-\sum_{(i,t)} C_{R_i}\delta(P(\theta_k|\mathbf{S}_{t}) < H_l),
\end{aligned}
\end{equation}
where a verification state $\mathbf{S}_t, t = 0,...,T$, is a vector of variables containing the results of all verification activities. 
Considering the verification model in Section~\ref{subsec:verification model} as an example, the verification state at time interval $t$ can be denoted as $\mathbf{S}_t = [A_1, A_2, A_3, A_4]$, where $A_i$ records the evidence of each activity node. Evidence can take on three values. If the node has not been verified, its value is 0. If the result of the VA (i.e., node evidence) is true (positive), its value is 1, and if the result is false (negative), its value is -1. $P(\theta_k|\mathbf{S}_t)$ is the conditional confidence level given the verification state $\mathbf{S}_t$ at time interval $t$. $\delta(\cdot)$ is an indicator function that captures the rework and deployment decision rules. Specifically for rework, $\delta(\cdot)$ equals 1 if $P(\theta_k|\mathbf{S}_{t})$ is lower than the threshold $H_l$; otherwise, its value is 0. For deployment, $\delta(\cdot)$ equals 1 if $P(\theta_k|\mathbf{S}_{T})$ is higher than the threshold $H_u$; otherwise, its value is 0.

The design of an optimal dynamic verification strategy $V_{opt}$ consists of finding a set of optimal verification activities that maximizes the expected value of the verification strategy at time $t=0,...,T-1$, considering that the expected value of all possible verification paths stemming from the verification strategy at the end $t = T$:
\begin{equation} \label{eq:2}
V_{opt} = \arg \max_{V_h}E_{\{\mathbf{S}_T\}}[ U(\mathbf{S}_T)|V_h=\{A_{i}|\mathbf{S}_t\} ],
\end{equation}
where a dynamic verification strategy $V_h$ consists of the activities $\{A_i\}$ for their corresponding verification states $\{\mathbf{S}_t\}$. 
Here, the notation $E_{\{\mathbf{S}_T\}}[\cdot]$ means taking expectation with respect to $\mathbf{S}_T$.

\subsection{Tree Search in Verification Tradespace}
\label{subsec:valuebaseddesign}
As discussed in Section~\ref{subsec:verification model}, a key challenge of designing verification strategies is the randomness of activity results, which means different results may lead to different verification paths. To account for this property of system verification, we formulate the verification tradespace as a directed tree space where each node of the tree represents one possible VA and the number of sub-branches of a node represents the number of results a VA can yield. For example, if all verification activities that could form a verification strategy have two possible results, the resulting tree space would become a binary directed tree. Hence, the maximum depth of the tree is given by the number of time intervals in which verification activities could be performed. Its width, however, is undetermined because the verification process could be stopped early in three situations. First, when reaching an intermediate time interval, a confidence level that is high enough to allow deployment of the system without requiring any further verification. Second, when reaching an intermediate time interval, a certain low confidence level could not be recovered, through rework and/or future VAs, into a sufficient confidence level that allows for an eventual deployment of the system. Third, it should be noted that, while it would be possible to opt to not execute any VA at a specific time interval (referred to as `NA'), for simplicity, it is assumed for this paper that once `NA’ is taken the verification process stops. Given this tree structure, all verification strategies can be organized as this type of directed trees in this study.

Instead of searching for the near-optimal tree in one step, the search for a dynamic verification strategy is focused on selecting a near-optimal VA that could maximize the expected value of all possible verification paths remaining for the rest of the time intervals, as suggested in prior work~\citep{xu2019concept}. However, as opposed to~\citep{xu2019concept}, we assume that the near-optimal VA is determined by choosing the root node of a foresight verification tree (FVT, denoted as $V^F_h$) that shares the same structure of the directed tree above. For example, considering the exemplar network in Fig.~\ref{fig:examplenetwork}, if the near-optimal FVT at the state $[0, 0, 0, 0]$ is the tree in Fig.~\ref{fig:SamplingIllus} (a), $A_2$ at the root node is selected as the near-optimal VA at $t=0$. After $A_2$ is implemented, if the state becomes $[0, 1, 0, 0]$, another FVT is explored in the same way to search for the next near-optimal activity. Similar to Eq.~\ref{eq:2}, optimality of a FVT $V^F_{opt}$ is then assessed using its expected value:

\begin{equation} \label{eq:3}
V^F_{opt} = \arg \max_{V^F_h}E_{\{\mathbf{S}_T\}}[U(\mathbf{S}_T)|V^F_h=\{A_{i}|\mathbf{S}_t\}].
\end{equation}

To calculate this expected value, $E(U)$, the posterior probabilities, $P_{p,q}$, of each branch $p$ of a path $q$ in $V^F_h$ are deduced using Bayesian inference on the BN model~\citep{koller2009probabilistic}. The probability of path $P_q$ can then be obtained by multiplying all probabilities of all branches along this path, $P_q = \prod_p P_{p,q}$. Thus, the expected value of this FVT $E(U)$ can be calculated as the weighted sum of the values of all paths: 
\begin{equation} \label{eq:4}
\begin{aligned}
E(U) &= E_{\{\mathbf{S}_{q,T}\}}(U(\mathbf{S}_{q,T}))= \sum_q P_q U(\mathbf{S}_{q,T}) \\
&= \sum_q (\prod_p P_{p,q}) U(\mathbf{S}_{q,T}),
\end{aligned}
\end{equation}
where $\{\mathbf{S}_{q,T}\}$ enumerates the verification states of all paths of a FVT at $t = T$.

Dynamic verification strategies are then evaluated by connecting the near-optimal activities sequentially as a hindsight verification tree (HVT, denoted as $V^H_{opt}$) after all possible states of a verification process are explored. The near-optimal activities of all possible states are connected according to their individual results. For example, if the first optimal action turns out to be $A_2$, all states following from $A_2$ onward, that is, $[0, 1, 0, 0]$ and $[0, -1, 0, 0]$, will be explored during the next time interval. For simplicity, the probabilities of all branches of $V^H_{opt}$ are deduced using Bayesian inference on the BN model~\citep{koller2009probabilistic}. So the distribution of these two states $P(A_2 = T/F)$ can be deduced from the BN model. Given the probabilities of all branches, the expected value of $V^H_{opt}$ can be calculated in the same way as Eq.~\ref{eq:4}.

\subsection{Proposed Parallel Tempering Approach}
\label{subsec:PTalgorithm}

\begin{figure}[htbp]
\vspace{-3mm}
\centerline{\includegraphics[width=8cm]{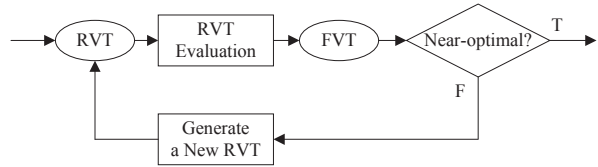}}
\caption{The Iterative Loop Method}
\label{fig:IterativeLoop}
\vspace{-3mm}
\end{figure}

\begin{figure*}
\centering
\subfigure[A RVT Example]{
    \begin{minipage}[b]{0.30\textwidth}
    \includegraphics[width=2.0in]{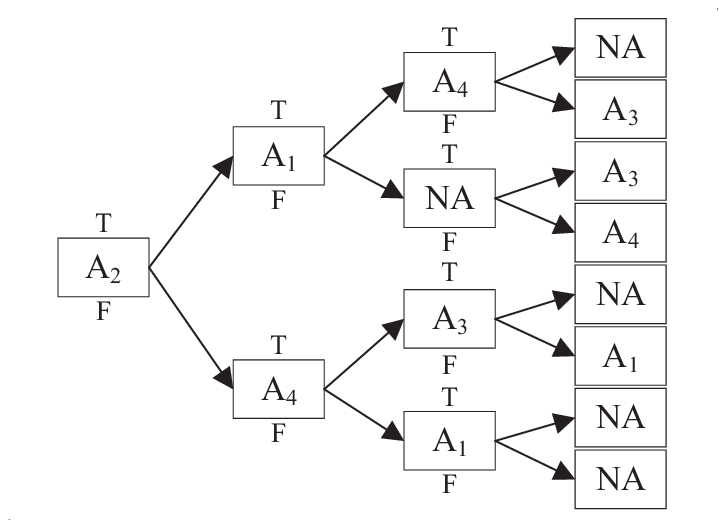}
    \end{minipage}
}
\subfigure[A FVT Example]{
    \begin{minipage}[b]{0.30\textwidth}
    \includegraphics[width=2.0in]{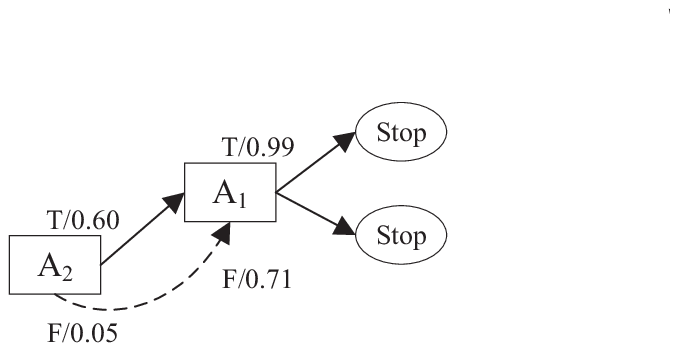}
    \end{minipage}
}
\subfigure[Sampling Weights]{
    \begin{minipage}[b]{0.30\textwidth}
    \includegraphics[width=2.0in]{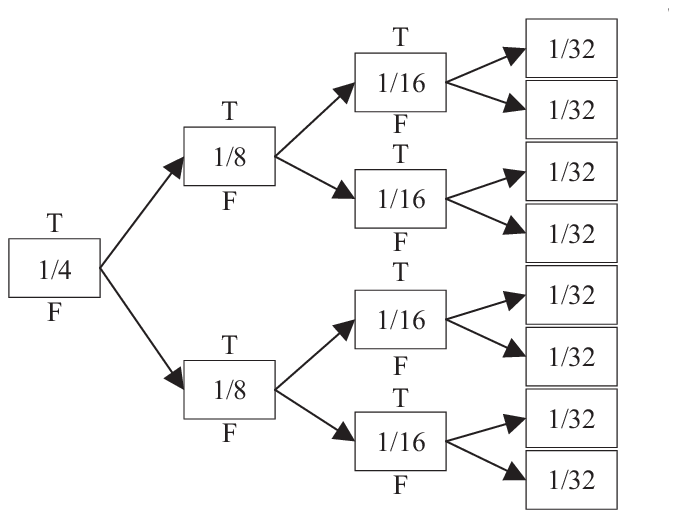}
    \end{minipage}
}
\caption{Illustration Example of the Iterative Loop Method. (a) A RVT $V^R_{h}$ is generated as a tree diagram; (b) A FVT $V^F_{h}$ is created from $V^R_{h}$; (c) All time intervals follow uniform distribution and the activities within each time interval share the same sampling weights.} 
\label{fig:SamplingIllus}
\vspace{-2mm}
\end{figure*}

The PT algorithm proposed in this paper leverages the standard PT framework, but it includes some modifications necessary to effectively address the specific characteristics of the system verification problem. The most important characteristic is that the execution of rework mechanisms influences the value of the verification strategy, but they remain uncertain when exploring the tradespace. To reduce the influence caused by this factor, the original MCMC of each replica was extended with an iterative loop to generate new samples, as shown in Fig~\ref{fig:IterativeLoop}. This loop first generated a raw verification tree (RVT), which is denoted as $V^R_h$. RVT is a tree diagram that has $n^{T-t}-1$ nodes, where there remains $T-t$ time intervals and each VA has $n$ results. All VAs that have not been implemented before $t$ can be selected as nodes of $V^R_h$. This RVT was then evaluated by going through all paths from its root node. If the attained confidence level is lower than the threshold $H_l$, a rework activity is triggered and the false result is corrected to a true result. If the confidence level reaches the threshold $H_u$ or `NA' is implemented, the verification process would be stopped. After the RVT evaluation is completed, all nodes of $V^R_h$ that have been visited form the corresponding FVT, $V^F_h$. In other words, given a $V^R_{h}$, its $V^F_{h}$ is created by pruning the non-visited branches with the rework rules in Section~\ref{subsec:performancemeasurement} and the early stopping rules in Section~\ref{subsec:valuebaseddesign}. For example, consider a RVT is generated as the tree in Fig.~\ref{fig:SamplingIllus} (a) and $H_l=0.2, H_u=0.95$. If the result of the first activity $A_2$ is false, a rework activity is triggered because $P(\theta_1) = 0.05 < 0.2$. So the following VA $A_4$ is pruned and only the following state $[0, 1, 0, 0]$ needs exploration. If the following $A_1$ is true, the process will stop because $P(\theta_1)=0.99 > 0.95$. If the result of $A_1$ is false, the process will also stop because the next activity is `NA'. The generated FVT $V^F_{h}$ is shown in Fig.~\ref{fig:SamplingIllus} (b). Finally, if $V^F_{h}$ is not near-optimal, another new RVT, $V^R_{h+1}$, will be generated for the next loop. Otherwise, $V^F_{h}$ is the exploration result.
 
Within this iterative loop method, the generation of RVTs can be realized in a similar way to the MCMC approach. In line with the MCMC idea that every sampling step is reversible (i.e., detailed balance condition), every new RVT $V^R_{h+1}$ would be generated from the previous RVT $V^R_{h}$ rather than from the previous FVT $V^F_{h}$. In this way, the invariance of the distribution of samples is ensured~\citep{tierney1994markov}. In addition, it is not possible to use the traditional statistic sampling method directly because, due to the tree structure of the samples, there is no specific distribution in the RVT tree space. Instead, new samples are generated in this study using the basic exchange and replacement rules. For simplicity, we assume that the exchange rule will be adopted with 80\% possibility while the replacement one will be adopted with 20\% possibility in practice. 

\begin{algorithm}
  \caption{Verification Activity Correction Algorithm}
  \label{alg:NoRepeat} 
  \begin{algorithmic}[1]
    \Inputs{$V^R_{h+1}=\{q_i\}$, $i=1,\ldots,I$, $X = \{A_1',A_2'\}.$}
    \Initialize{A position set: $Y=\{$\o$\}.$}
    \State Add the positions of $A_1'$ and $A_2'$ to $Y$.
    \While{$r = True$}
      \State $r = False$.
      \For{$i = 1$ to $I$}
        \If{An activity $A_j'\in X$ is executed twice in $q_i$}
            \State \parbox[t]{185pt}{Identify the position $(i,j)$ from the two positions that is not within $Y$.\strut}
            \State \parbox[t]{185pt}{Add the position $(i,j)$ to $Y$.\strut}
            \State \parbox[t]{185pt}{Replace $A_j'$ at $(i,j)$ with another activity in $X$.\strut}
            \State $r = True$.
        \EndIf
      \EndFor
    \EndWhile
  \end{algorithmic}
\end{algorithm}

To be more specific, the exchange rule is used to randomly select two verification activities $\{A'_1, A'_2\}$ from $V^R_{h}$ and switch their node positions to generate new samples. In particular, there is a restriction that each VA can be executed only once along each verification path $q_i$. Taking the RVT in Fig.~\ref{fig:SamplingIllus} (a) as an example, if the activities $\{A_2,A_1\}$ in the first path $\{A_2, A_1, A_4, NA\}$ are switched, there is an activity conflict in the last two paths $q_7/q_8 = \{A_1, A_4, A_1, NA\}$ because $A_1$ is executed twice. The second $A_1$ in $q_7/q_8$ can be replaced with $A_2$ to correct this conflict. Because this kind of activity conflict may happen in an unpredictable way, we examine all paths and correct conflicting activities iteratively until there are no activity conflicts. A verification activity correction algorithm is added to correct all conflicting activities along all paths $\{q_i\}$ of $V^R_{h+1}$ after the exchange, as shown in Algorithm~\ref{alg:NoRepeat}. In addition, the replacement rule is used to replace a randomly selected VA, $A_s$, with another target activity to generate new samples. The target VA is sampled from all candidate activities that do not appear along the paths of the activity $A_s$. `NA' is also included as a candidate VA. For simplicity, candidate verification activities are randomly chosen according to a uniform distribution.

As an acceleration technique of the two rules, all activities in the previous RVT $V^R_{h}$ are assigned with weights according to their node positions. That is, the sampling weights of all time intervals, as well as those of all verification activities within one time interval, follow the uniform distribution. This is done by setting the probability of importance (i.e., weight) for each VA $W_{i}$ as the reverse function of the number of remaining $T-t$ time intervals and the number of branches $N_t$ at its time interval $t$. The formula is $W_i=\frac{1}{N_t \cdot (T-t)}$. One illustration example is shown in Fig.~\ref{fig:SamplingIllus} (c). Given such weights of all node positions, a new sample $V^R_{h+1}$ can be generated from the original one $V^R_{h}$ with the two rules above. 

Another modification of the standard PT algorithm is the specification of parameters. First, different from the minimization problem of the standard PT method, the target function used in this study is to maximize the expected value of replicas. So the acceptance probability is changed to $P_e = min(1,exp(-\Delta \beta \Delta E))$, where $\Delta \beta = \frac{1}{\Psi_m}-\frac{1}{\Psi_{m+1}}$, $\Delta E=E_m - E_{m+1}$, and $E_m$ is the expected value of $V^F_h(\Psi_m)$ in the replica $\Omega(\Psi_m)$. Second, the temperatures $\{\Psi_m\}$, as major hyperparameters of the PT algorithm, have a direct impact on the acceptance probability. Considering the rules of thumb presented in Subsection~\ref{subsec:paralleltempering}, three conditions are identified to determine the temperatures. 
The first condition is that $P_e$ should be larger than some threshold $C_1$ for the pair of replicas with the two highest temperatures $\{\Omega(\Psi_{M-1}),\Omega(\Psi_M)\}$, even when $\Delta E$ is maximum. So the constraint about $\Delta \beta$ can be deduced as:
\begin{equation} \label{eq:301}
\begin{aligned}
exp(-\Delta \beta \Delta E_{max}) &> C_1, \\
\Delta \beta &> -\frac{log(C_1)}{\Delta E_{max}}.
\end{aligned}
\end{equation}
The second condition is that $P_e$ should be smaller than some threshold $C_2$ for the pair with the two lowest temperatures $\{\Omega(\Psi_1),\Omega(\Psi_2)\}$ as long as $\Delta E$ is larger than a threshold $\Delta E_{thres}$. So another constraint about $\Delta \beta$ can be deduced as:
\begin{equation} \label{eq:302}
\begin{aligned}
exp(-\Delta \beta \Delta E_{thres}) &< C_2, \\
\Delta \beta &< -\frac{log(C_2)}{\Delta E_{thres}}.
\end{aligned}
\end{equation}
Then the range of $\Delta \beta$ is $[-\frac{log(C_1)}{\Delta E_{max}}, -\frac{log(C_2)}{\Delta E_{thres}}]$. Next, following the analytical study in~\citep{kofke2002acceptance}, we assumed there is a constant ratio value $C_3=\frac{\Psi_{m+1}}{\Psi_m}$ between all pairs of temperatures for simplicity. The set of temperatures can be calculated with these three conditions. While the specific values of temperatures depend on the initial conditions of the experiment, their calculation is discussed in Section~\ref{subsec:experimentsetup}.

Finally, a convergence rule is proposed to obtain a satisfactory FVT solution with limited computational resources. In the standard PT algorithm, the total number of swaps is determined first, which makes it hard to compare benchmark methods. Thus, the convergence of tree search is determined according to the duration of near-optimal solutions. As all replicas of the verification process are repeatedly sampled over time, they are divided into a series of non-overlapping periods, which we term the window period. The length of each window period is set as $N_{it}$. Within each period, the optimal FVT sample, $V^F_{opt}$, can be found from all replicas. The basic idea is that if $V^F_{opt}$ remains the best alternative after a certain number of replica iterations (named Convergence Length $L$), we treat it as the near-optimal one. Determining the specific value of $L$ requires experimental tests because its value depends on the specific context, which will be discussed in Section~\ref{subsec:Discussion}. Note, once $L$ is determined, it is unnecessary to specify $N_s$ because both function as constraints on the total length of the replica iterations. 

In summary, compared with Algorithm~\ref{alg:ClassicPT}, the proposed PT approach presents three main modifications: the iterative loop method, the specification of parameters, and the convergence rule. The complete PT algorithm used in this paper is shown in Algorithm~\ref{alg:PT}.

\begin{algorithm}
  \caption{Proposed PT Algorithm}
  \label{alg:PT} 
  \begin{algorithmic}[1]
    \Inputs{$N_{it}, \{\Psi_m\}, m=1,\ldots,M.$}
    \Initialize{$\{\Omega(\Psi_m)\}:\Omega(\Psi_m) = V^R_0(\Psi_m)$.}
    \While{True}
      \For{m = 1 to M}
        \State \parbox[t]{200pt}{Apply the iterative loop method to $\Omega(\Psi_m)$ for $N_{it}$ iterations.\strut}
      \EndFor
      \For{m = 1 to M-1}
        \State \parbox[t]{200pt}{Swap $V^R_h(\Psi_m)$ with $V^R_h(\Psi_{m+1})$ with the probability $p=min(1,exp(-\Delta\beta\Delta E))$.\strut}
      \EndFor
      \State \parbox[t]{215pt}{Search for the best sample $V^F_{opt}$ from $\{\Omega(\Psi_m)\}$.\strut}
    \If{$V^F_{opt}$ meetsthe proposed convergence rule}
      \State Stop.
    \EndIf
    \EndWhile
  \end{algorithmic}
\end{algorithm}

\section{Experimental Design}
\label{sec:experiment}
In this section, we apply the proposed methodology to design a dynamic verification strategy for an optical instrument in a satellite. The performance of the proposed method is assessed in sixteen cases having different complexity.

\begin{figure}[htbp]
\vspace{-3mm}
\centerline{\includegraphics[width=5.5cm]{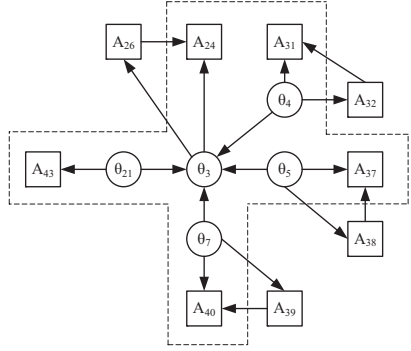}}
\caption{Small Network (within the dashed line) and Medium Network (i.e., the whole graph)}
\label{fig:smallmidnet}
\vspace{-3mm}
\end{figure}

\begin{figure*}[htbp]
\vspace{-5mm}
\centerline{\includegraphics[width=15cm]{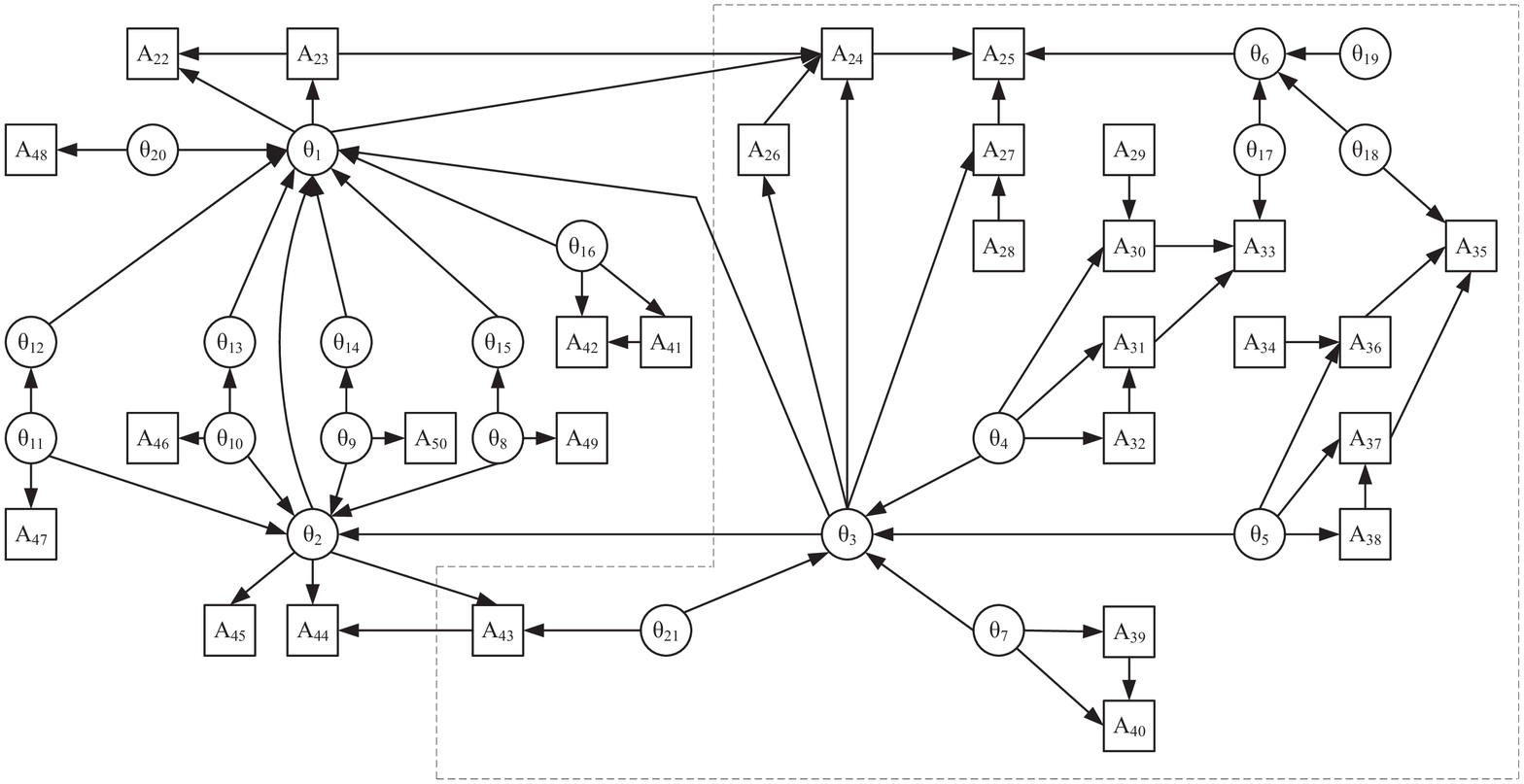}}
\caption{Large Network (within the dashed line) and Full Network (i.e., the whole graph)}
\label{fig:fullnetwork}
\vspace{-5mm}
\end{figure*}

\subsection{Experimental Setup}
\label{subsec:experimentsetup}
We use an optical instrument as the system model and a notional set of possible verification activities presented in\citep{Salado2019} as the test case for this study. The engineered system and its possible verification activities are represented, as shown in Fig.~\ref{fig:fullnetwork}, as a BN where system parameters are shown as circle nodes and candidate verification activities are denoted as square nodes. The full network contains three components that contribute to the field of view ($\theta_1$), the modular transfer function ($\theta_3$), and the system degradation ($\theta_6$). The definition of each node is given in\citep{Salado2019} and is not reproduced here, since they do not affect the results of this paper. Each node is characterized in this paper with its own conditional probability table (CPT), which is provided as a digital file. Specific values are synthetic and have been generated using the Noisy-or model~\citep{pearl2014probabilistic}, which takes into account the physical meaning of the different modes when estimating their mutual effects for reasonability of the data. 

In this experiment, we assume that system revenue is driven by system parameter $\theta_3$. Hence, $\theta_3$ is set as the single target parameter. The number of time intervals is set as $5$, as it provides sufficient complexity to demonstrate the performance of PTA without requiring extensive computational effort. Four types of rework rules are used in the study to explore different interpretations of the rework triggering mechanism. They are modeled as different values of $H_l$ at each time interval and are referred to as `Low', `Low-high', `High-low', and `High'. The specific values of $H_l$ are shown in Table~\ref{tab:LowerThres}. A low $H_l$ at some time interval models a situation in which immediate rework may not be necessary. In contrast, a high $H_l$ models a situation in which rework must be performed straight away. The threshold for the system deployment rule, $H_u$, is set as 0.95.  

\begin{table}[]
\centering
\caption{Lower Thresholds of Different Rework Rules}
\label{tab:LowerThres}
\begin{tabular}{|c|ccccc|}
\hline
Time Interval (t) & $0$ & $1$ & $2$ & $3$ & $4$ \\ \hline
\hline
Low & 0.2 & 0.2 & 0.2 & 0.2 & 0.2 \\
Low-high & 0.2 & 0.3 & 0.575 & 0.85 & 0.95 \\
High-low & 0.95 & 0.85 & 0.575 & 0.3 & 0.2 \\
High & 0.95 & 0.95 & 0.95 & 0.95 & 0.95 \\ \hline
\end{tabular}
\vspace{-5mm}
\end{table}

Four networks of different sizes are used to explore the scalability of the approach with problem size. First, we apply the PTA method to a small network, consisting only of system attributes and verification activities related to $\theta_3$. This is outlined by the dashed line in Fig.~\ref{fig:smallmidnet}. Second, we apply the proposed PTA to a medium network whose nodes are closely related to $\theta_3$, as shown in Fig~\ref{fig:smallmidnet}. Third, we reduce the scope of the proposed PTA to a large network whose nodes are related to $\theta_3$ and $\theta_6$. It is outlined by the dashed line in Fig.~\ref{fig:fullnetwork}. Finally, we apply the proposed PTA to the full network in Fig.~\ref{fig:fullnetwork}. These four types of networks share the same parameter $\theta_3$ as the target node. They have scaling relationships from some closest nodes to all connected nodes. To ensure that the four networks are comparable, we set the joint distributions of all subnetworks, including small, medium, and large networks, as the marginal distribution of the full network. That is, the probabilistic relationships defined for the full network are completely reserved in all four types of networks.

Cost data have also been synthetically generated in thousand dollar units (\$1,000). The revenue $B_k$ has been set to 20,000 so that it provides a balance when choosing verification activities. The execution costs of the different VAs, as well as the corresponding rework costs, are provided in Table 2. The execution costs have been generated from a range $[250, 1000]$. Specific values have been defined according to the type of VA defined in~\citep{Salado2019}. In particular, VAs directly associated with system parameters $\theta_1$, $\theta_2$, $\theta_3$, and $\theta_6$ have been considered to be more expensive than the rest. 

Rework costs associated with each node $A_i$ have been designed by considering two factors, as defined in~\citep{Salado2019}: the development phase in which the VA is executed (e.g., Preliminary Design Review (PDR), Critical Design Review (CDR), etc.) and the type of system parameter (including verification model) the VA verifies. In general, it is assumed that the later the rework is executed in the system development, the higher the rework cost is~\citep{blanchard1990systems}. This rule of thumb is also incorporated as a penalty factor to promote rework happening as early as possible. The rework penalty factor was specifically defined as the multiplication coefficient $[1, 1.11, 1.22, 1.36, 1.5]$, where each element in the vector corresponds to increasing time intervals. For example, executing rework after $A_{23}$ at the first time interval $t=0$ costs $740$, while executing the same rework after $A_{23}$ occurs at $t=4$ would cost $740*1.5 = 1110$. To simplify the experiment, the values of all cost items are assumed to be fixed and known beforehand.

The specification of parameters follows the rules presented in Section~\ref{subsec:PTalgorithm}. The temperatures are determined according to the three conditions described earlier. Given the cost values defined in Table~\ref{tab:costtable}, the range of $\Delta E$ within $5$ time intervals is within $[-3.8*10^{5},3.8*10^{5}]$. So $\Delta E_{max} = 3.8*10^{5}$. We also assign the constants with values $C_1 = 0.05, C_2 = 0.05, C_3 = 2, \Delta E_{thres} = 100$. Then the range of $\Delta \beta$ is calculated as $[0.79*10^{-5}, 0.03]$. The final temperatures has been set as 
\begin{equation} \label{eq:5}
\begin{aligned}
\{\Psi_m\} = &\{10, 20, 39, 78, 156, 312, 625, 1250, 2500, \\
&5000, 10000, 20000, 40000, 80000, 160000\}
\end{aligned}
\end{equation}
to cover the range of $\Delta \beta$. The convergence length $L$ is set at $1,000$ iterations, as discussed in Subsection~\ref{subsec:Discussion}. We also set $N_{it} = 50$ (i.e., the length of the window period is 50 iterations), as it does not yield any sensitivity for fixed convergence length. 

Finally, the PTA has been implemented with 15 cores (for each core, CPU: Intel E5-2683V4 2.1GHz, Memory: 4GB) in the parallel computing environment provided by Advanced Research Computing at Virginia Tech.

\subsection{Experimental Method}
\label{subsec:ExperimentalMethod}
The proposed PTA in this experiment is compared against several benchmark methods, including the fixed path method (FP), the basic Monte Carlo method (MC), the dynamic Monte Carlo method (DMC), and the static FVT approach (SFVT). The FP represents the approach commonly used in verification engineering practice. In essence, a set of VAs is defined at the beginning of system development, which are strictly conducted regardless of their results. While the selection of the specific set of activities (and path) is performed following industry standards or subject matter expert experience~\citep{Engel2010}, an optimal path found via full enumeration is used in this case. Hence, the FP benchmark used in this paper represents a best-case scenario of industry practice. The MC is based on the random generation of solution trees, which are compared in terms of their expected values, with the best one among the set being chosen as a static verification strategy. The DMC combines the MC method and dynamic design, such that, for each possible verification state of a verification process, MC is used to identify a near-optimal VA. Finally, to illustrate the effect of dynamic design, we consider the SFVT, which uses the near-optimal FVT generated at $T=0$ as the static verification strategy. 

To make a fair comparison, we grouped the proposed PTA method and the DMC method together and set the rest of the methods as a separate group to show the effect of dynamic design. As to run time, we apply the same convergence rule (i.e., $L = 1,000$) and the parallel computing environment (15 cores) to all these methods under the 16 cases. The wall clock time (i.e., elapsed real time) is used to compare computational efficiency.

\begin{table*}[]
\centering
\caption{Verification Activity Execution Costs and Rework Costs (Unit: \$1,000)}
\label{tab:costtable}
\begin{tabular}{|c|cccccccccc|}
\hline
Verification Activity &  & $A_{22}$ & $A_{23}$ & $A_{24}$ & $A_{25}$ & $A_{26}$ & $A_{27}$ & $A_{28}$ & $A_{29}$ & $A_{30}$ \\ \hline
Activity Cost ($C_A$) &  & 350 & 800 & 350 & 250 & 800 & 350 & 350 & 550 & 450 \\
Rework Cost ($C_R$) &  & 39,010 & 740 & 36,620 & 38,430 & 5,160 & 37,550 & 30,970 & 8,310 & 7,030 \\ \hline
\hline
Verification Activity & $A_{31}$ & $A_{32}$ & $A_{33}$ & $A_{34}$ & $A_{35}$ & $A_{36}$ & $A_{37}$ & $A_{38}$ & $A_{39}$ & $A_{40}$ \\ \hline
Activity Cost ($C_A$) & 300 & 250 & 700 & 250 & 700 & 450 & 300 & 350 & 350 & 550 \\
Rework Cost ($C_R$) & 7,880 & 1,860 & 8,180 & 6,200 & 8,070 & 6,020 & 7,800 & 1,490 & 770 & 7,910 \\ \hline
\hline
Verification Activity & $A_{41}$ & $A_{42}$ & $A_{43}$ & $A_{44}$ & $A_{45}$ & $A_{46}$ & $A_{47}$ & $A_{48}$ & $A_{49}$ & $A_{50}$ \\ \hline
Activity Cost ($C_A$) & 1000 & 450 & 450 & 950 & 950 & 250 & 250 & 400 & 850 & 250 \\
Rework Cost ($C_R$) & 740 & 8,020 & 1,700 & 1,470 & 1,270 & 1,160 & 1,600 & 1,330 & 1,010 & 1,220 \\ \hline
\end{tabular}
\end{table*}

\subsection{Experimental Results}
\label{subsec:ExpResults}
\begin{figure}
\vspace{-1mm}
\centering
\subfigure[The 1st FVT]{
\begin{minipage}[b]{0.5\textwidth}
  \includegraphics[width=2.4in]{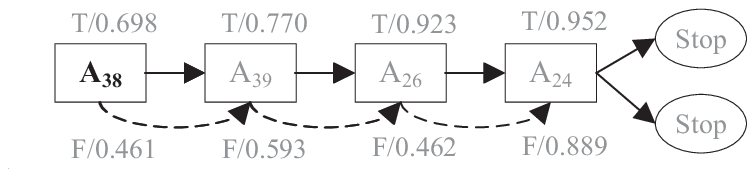}
  \end{minipage}
}
\subfigure[The 2nd FVT]{
\begin{minipage}[b]{0.5\textwidth}
  \includegraphics[width=2.4in]{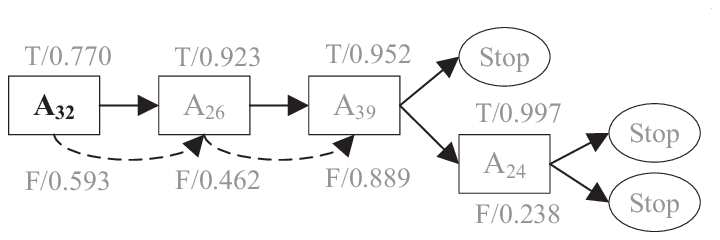}
  \end{minipage}
}
\subfigure[The 3rd FVT]{
\begin{minipage}[b]{0.5\textwidth}
  \includegraphics[width=2.4in]{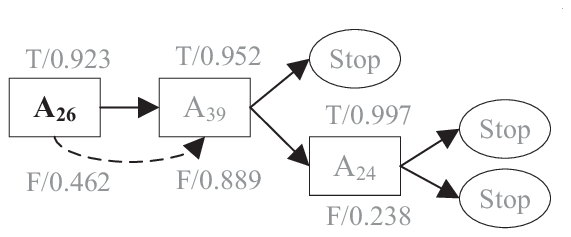}
  \end{minipage}
}
\subfigure[The 4th FVT]{
\begin{minipage}[b]{0.5\textwidth}
  \includegraphics[width=2.4in]{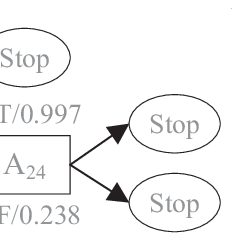}
  \end{minipage}
}
\subfigure[The 5th FVT]{
\begin{minipage}[b]{0.5\textwidth}
  \includegraphics[width=2.4in]{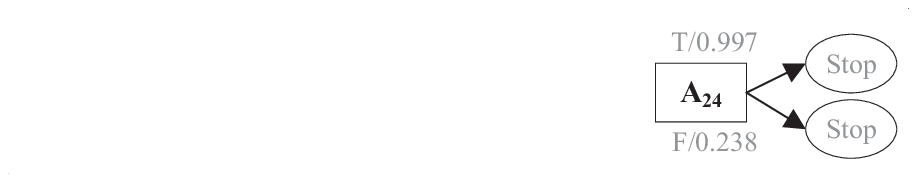}
  \end{minipage}
}
\caption{Dynamic Generation of FVTs} 
\label{fig:DPTIllus}
\vspace{-5mm}
\end{figure}

\begin{figure*}
\vspace{-2mm}
\centering
\subfigure[Small / Low]{
    \begin{minipage}[b]{0.13\textwidth}
    \includegraphics[width=1.2in]{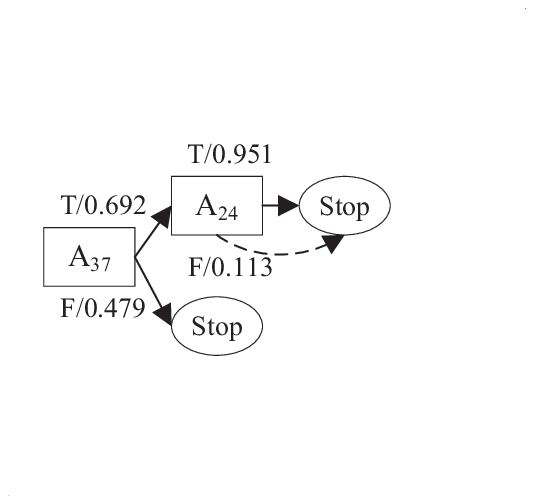}
    \end{minipage}
}
\subfigure[Medium / Low]{
    \begin{minipage}[b]{0.27\textwidth}
    \includegraphics[width=1.9in]{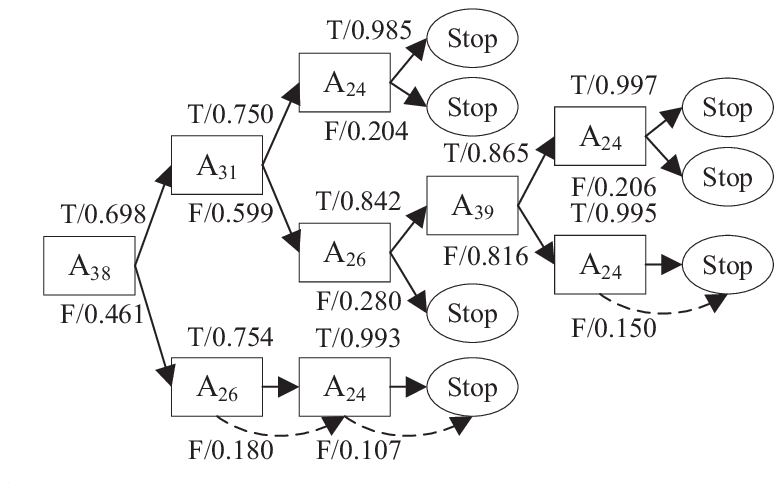}
    \end{minipage}
}
\subfigure[Large / Low]{
    \begin{minipage}[b]{0.27\textwidth}
    \includegraphics[width=1.9in]{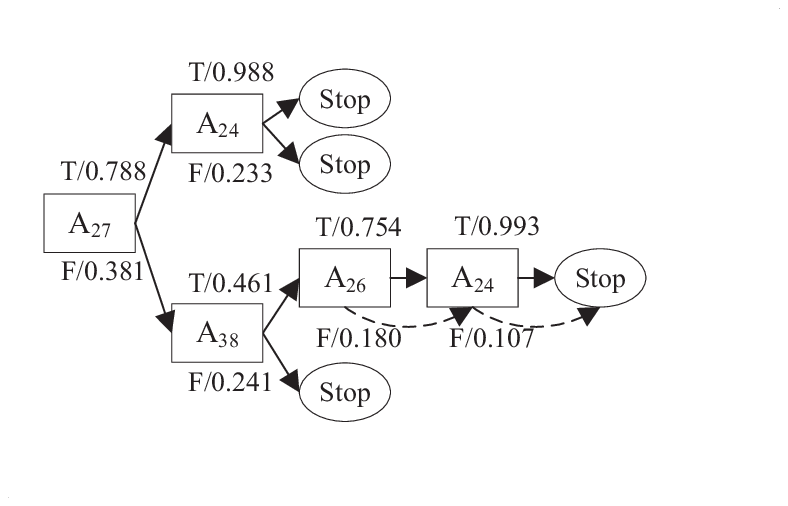}
    \end{minipage}
}
\subfigure[Full / Low]{
    \begin{minipage}[b]{0.27\textwidth}
    \includegraphics[width=1.9in]{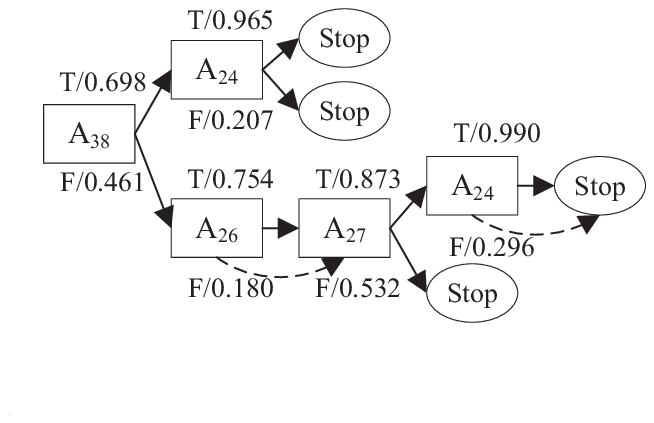}
    \end{minipage}
}

\subfigure[Small / Low-high]{
    \begin{minipage}[b]{0.13\textwidth}
    \includegraphics[width=1.2in]{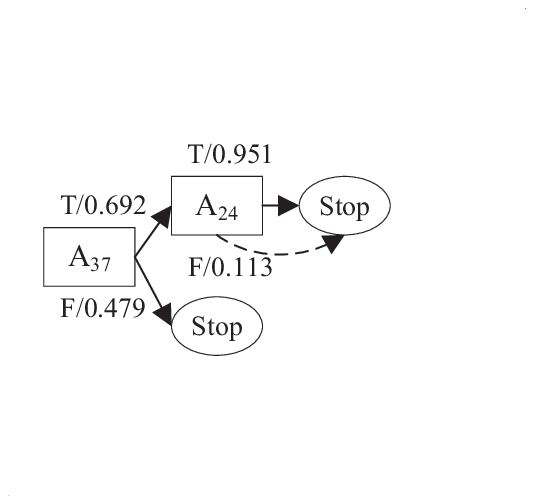}
    \end{minipage}
}
\subfigure[Medium / Low-high]{
    \begin{minipage}[b]{0.27\textwidth}
    \includegraphics[width=1.9in]{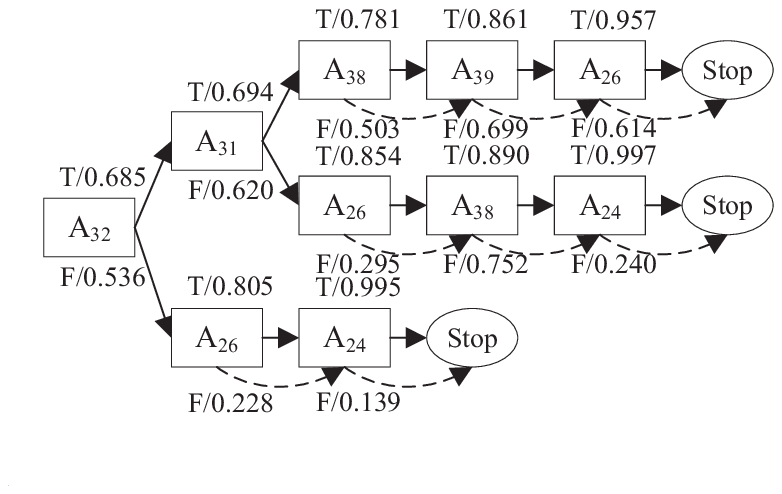}
    \end{minipage}
}
\subfigure[Large / Low-high]{
    \begin{minipage}[b]{0.27\textwidth}
    \includegraphics[width=1.9in]{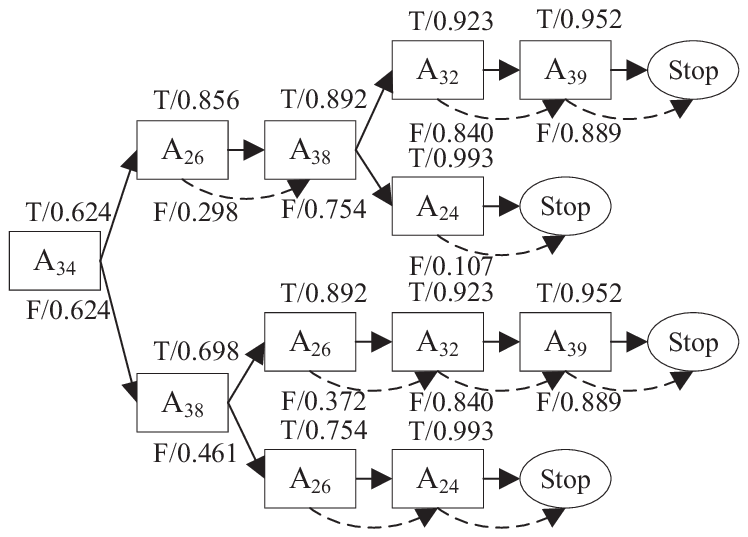}
    \end{minipage}
}
\subfigure[Full / Low-high]{
    \begin{minipage}[b]{0.27\textwidth}
    \includegraphics[width=1.9in]{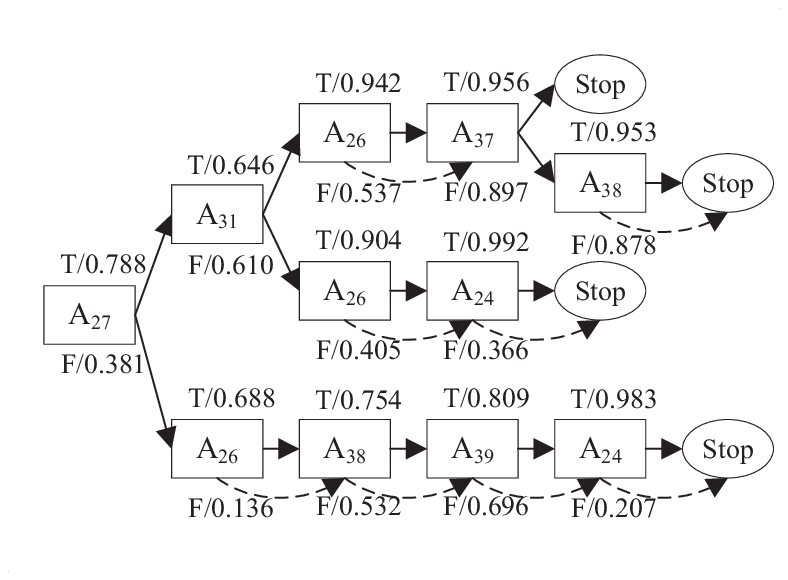}
    \end{minipage}
}

\subfigure[Small / High-low]{
    \begin{minipage}[b]{0.13\textwidth}
    \includegraphics[width=1.2in]{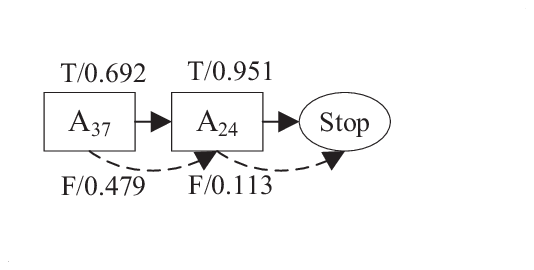}
    \end{minipage}
}
\subfigure[Medium / High-low]{
    \begin{minipage}[b]{0.27\textwidth}
    \includegraphics[width=1.9in]{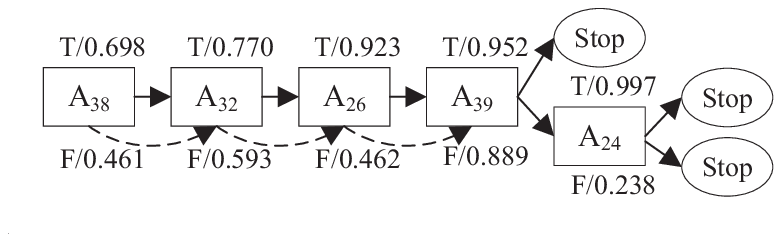}
    \end{minipage}
}
\subfigure[Large / High-low]{
    \begin{minipage}[b]{0.27\textwidth}
    \includegraphics[width=1.9in]{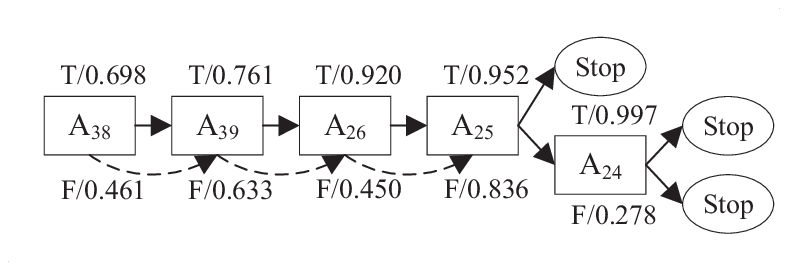}
    \end{minipage}
}
\subfigure[Full / High-low]{
    \begin{minipage}[b]{0.27\textwidth}
    \includegraphics[width=1.9in]{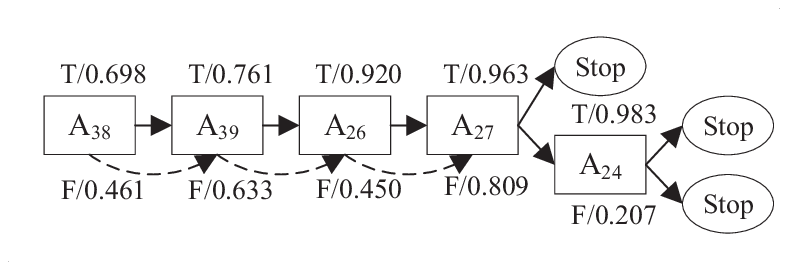}
    \end{minipage}
}

\subfigure[Small / High]{
    \begin{minipage}[b]{0.13\textwidth}
    \includegraphics[width=1.2in]{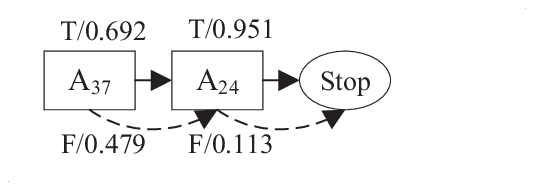}
    \end{minipage}
}
\subfigure[Medium / High]{
    \begin{minipage}[b]{0.27\textwidth}
    \includegraphics[width=1.9in]{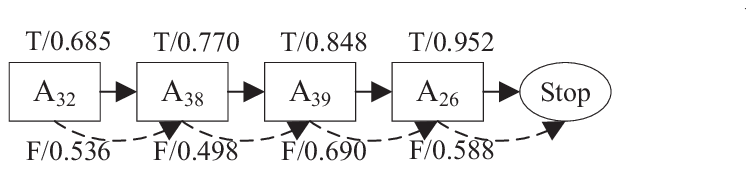}
    \end{minipage}
}
\subfigure[Large / High]{
    \begin{minipage}[b]{0.27\textwidth}
    \includegraphics[width=1.9in]{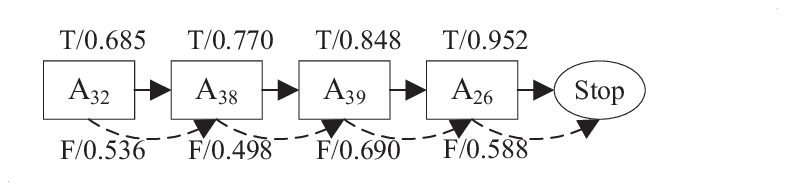}
    \end{minipage}
}
\subfigure[Full / High]{
    \begin{minipage}[b]{0.27\textwidth}
    \includegraphics[width=1.9in]{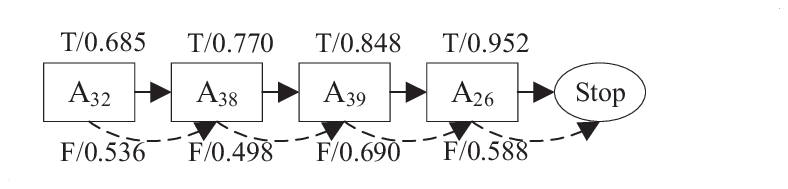}
    \end{minipage}
}

\caption{HVTs of the 16 Cases}
\label{fig:HVTs}
\vspace{-5mm}
\end{figure*}

We applied the proposed PTA to all 16 cases that are combinations between four networks and four rework rules in this experiment. As shown in Fig.~\ref{fig:HVTs}, the sixteen HVTs are generated by the algorithm and can be read as follows: (1) These HVTs summarize the root nodes of the near-optimal FVT at each verification state. For example, in Case (j), which has a medium network and high-low rework rule, a FVT is generated first at $t=0$, as shown in Fig.~\ref{fig:DPTIllus} (a). $A_{38}$ is chosen as the first near-optimal VA. If the result of $A_{38}$ is true, another FVT is generated at $t=1$, as shown in Fig.~\ref{fig:DPTIllus} (b). $A_{32}$ is recommended as the next near-optimal VA. The same reasoning applies throughout the five time intervals. All near-optimal activities are connected as the HVT in Fig.~\ref{fig:HVTs} (j). (2) The posterior confidence on the correct functioning of the system, i.e., that of the target node $\theta_3$, has been labelled next to each activity result (T or F) in Fig.~\ref{fig:HVTs}. As expected, the posterior confidence is shaped by verification results~\citep{salado2019capturing}. If $P(\theta_3)$ is lower than the lower threshold $H_l$, a rework activity is triggered, which is represented as a dashed curve. (3) Stop endpoints indicate the early stopping rules in Section~\ref{subsec:valuebaseddesign}. If $P(\theta_3)$ is larger than $H_u$, the system can be deployed and there is a stop endpoint. Otherwise, a stop endpoint indicates the near-optimal activity of a FVT is `Stop'. That is, the certain low confidence level could not be recovered into a high confidence level through rework and/or future VAs.

\begin{figure*}
\vspace{-3mm}
\centering
\subfigure[Small / Low]{
    \begin{minipage}[b]{0.23\textwidth}
    \includegraphics[width=1.5in]{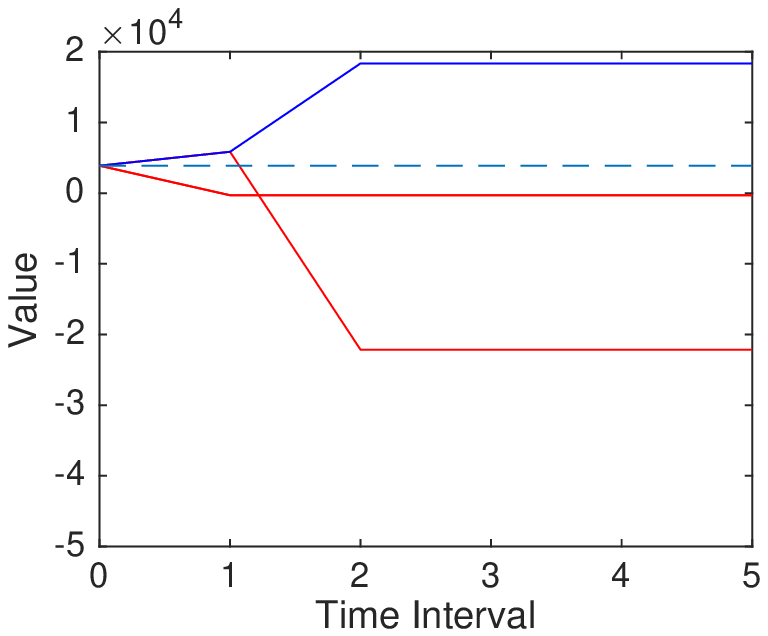}
    \end{minipage}
}
\subfigure[Medium / Low]{
    \begin{minipage}[b]{0.23\textwidth}
    \includegraphics[width=1.5in]{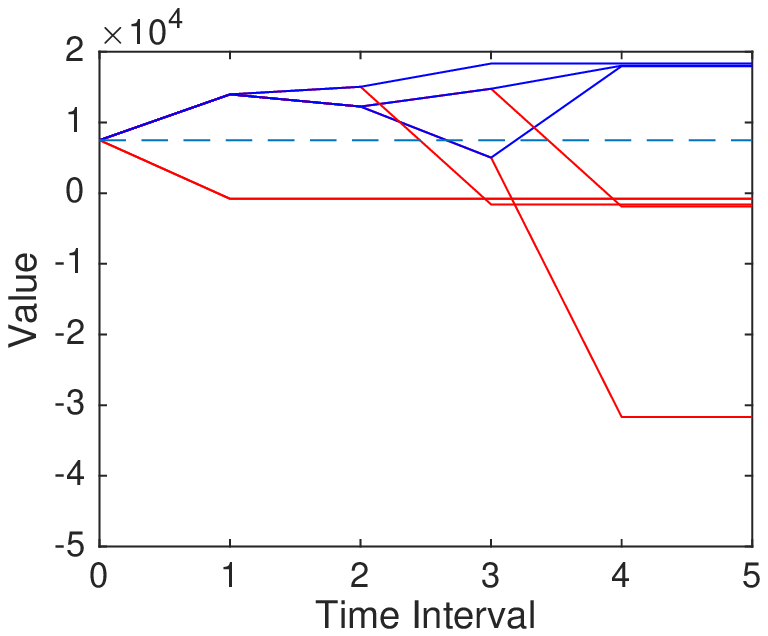}
    \end{minipage}
}
\subfigure[Large / Low]{
    \begin{minipage}[b]{0.23\textwidth}
    \includegraphics[width=1.5in]{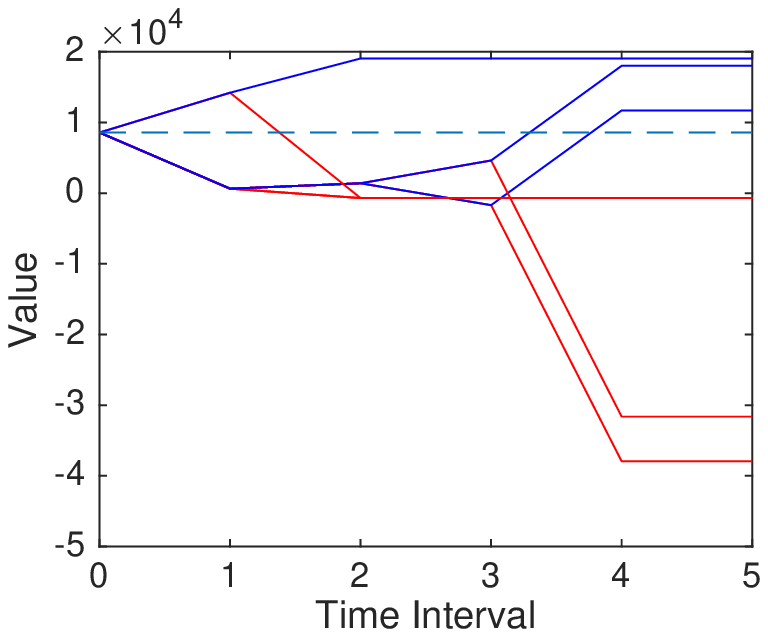}
    \end{minipage}
}
\subfigure[Full / Low]{
    \begin{minipage}[b]{0.23\textwidth}
    \includegraphics[width=1.5in]{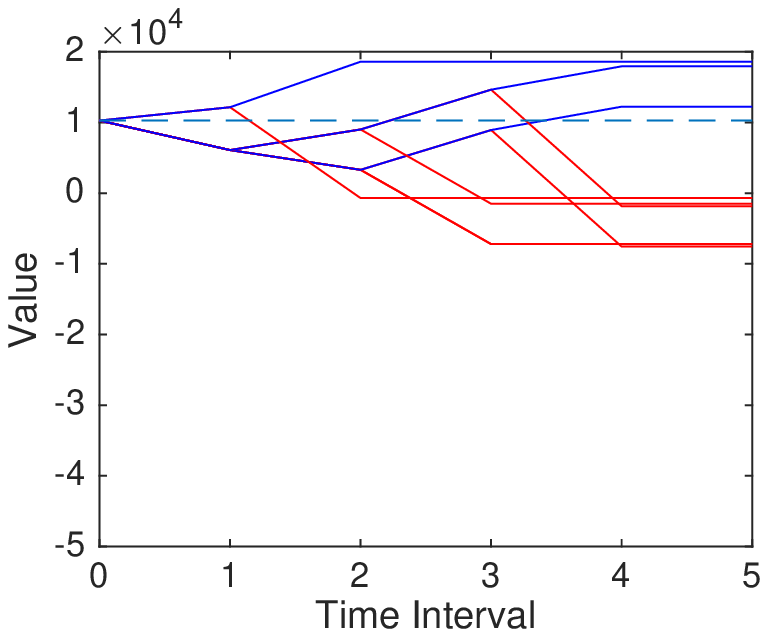}
    \end{minipage}
}

\subfigure[Small / Low-high]{
    \begin{minipage}[b]{0.23\textwidth}
    \includegraphics[width=1.5in]{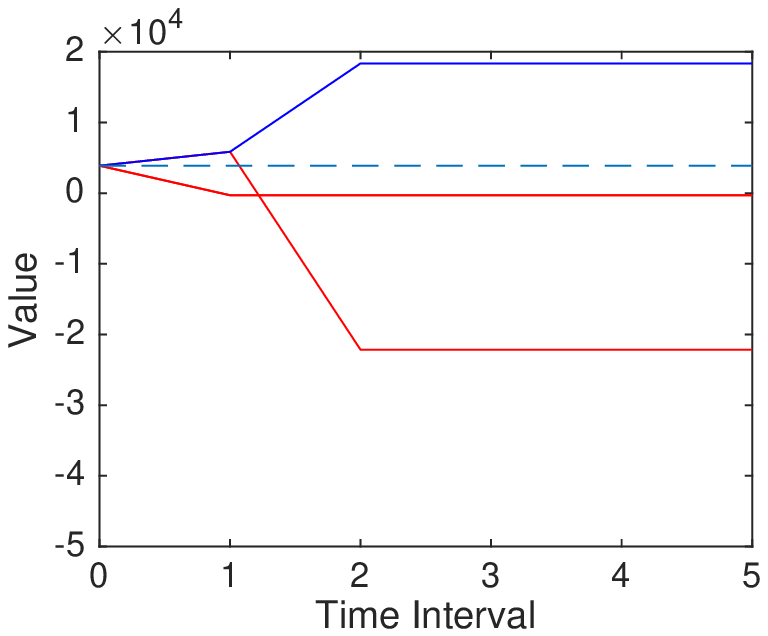}
    \end{minipage}
}
\subfigure[Medium / Low-high]{
    \begin{minipage}[b]{0.23\textwidth}
    \includegraphics[width=1.5in]{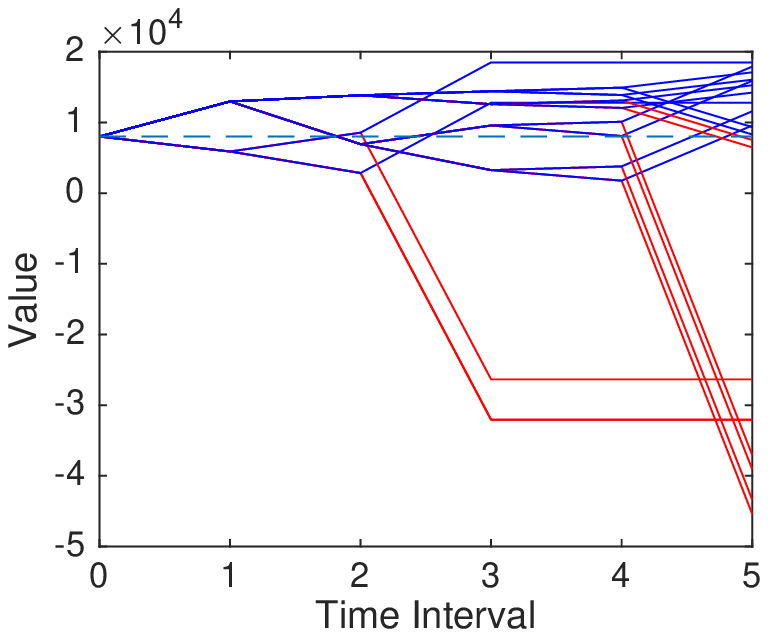}
    \end{minipage}
}
\subfigure[Large / Low-high]{
    \begin{minipage}[b]{0.23\textwidth}
    \includegraphics[width=1.5in]{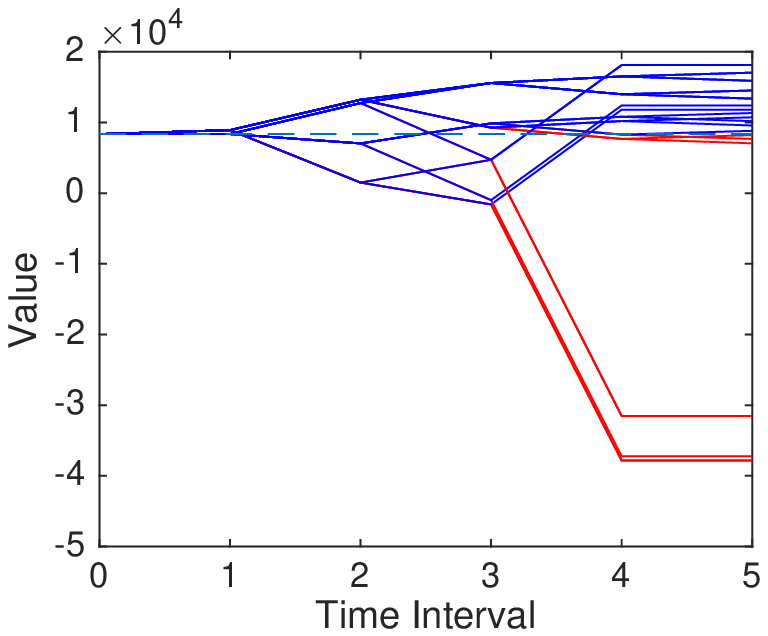}
    \end{minipage}
}
\subfigure[Full / Low-high]{
    \begin{minipage}[b]{0.23\textwidth}
    \includegraphics[width=1.5in]{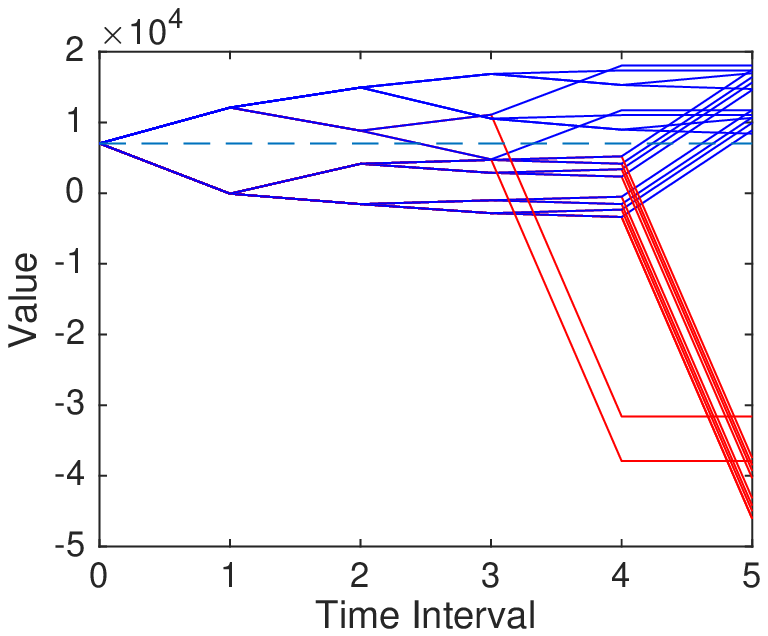}
    \end{minipage}
}

\subfigure[Small / High-low]{
    \begin{minipage}[b]{0.23\textwidth}
    \includegraphics[width=1.5in]{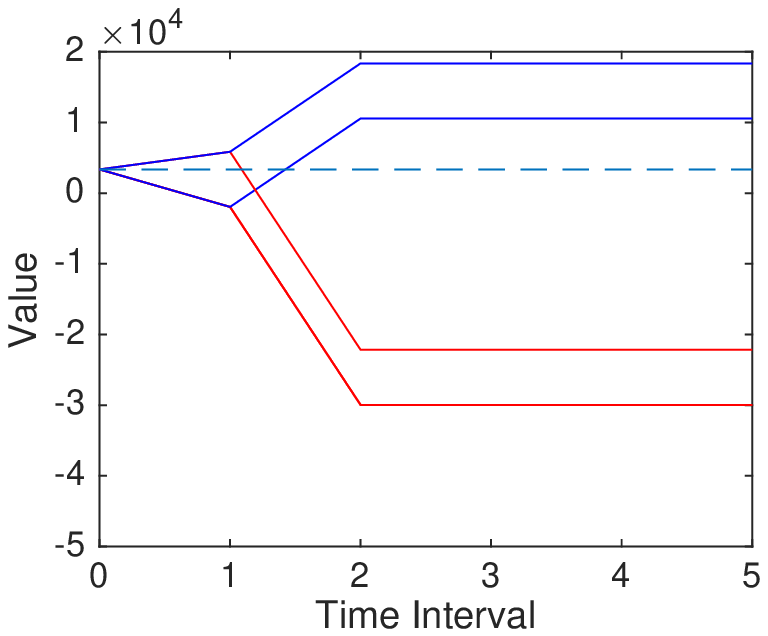}
    \end{minipage}
}
\subfigure[Medium / High-low]{
    \begin{minipage}[b]{0.23\textwidth}
    \includegraphics[width=1.5in]{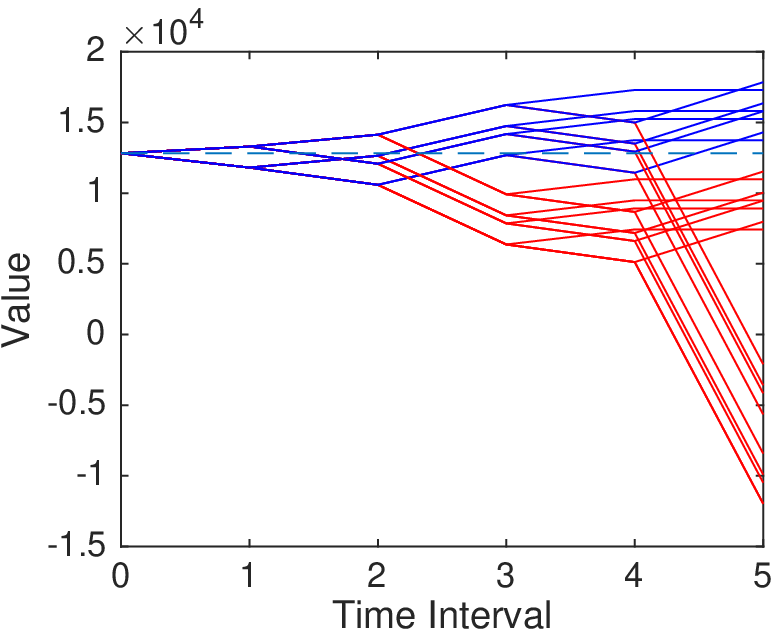}
    \end{minipage}
}
\subfigure[Large / High-low]{
    \begin{minipage}[b]{0.23\textwidth}
    \includegraphics[width=1.5in]{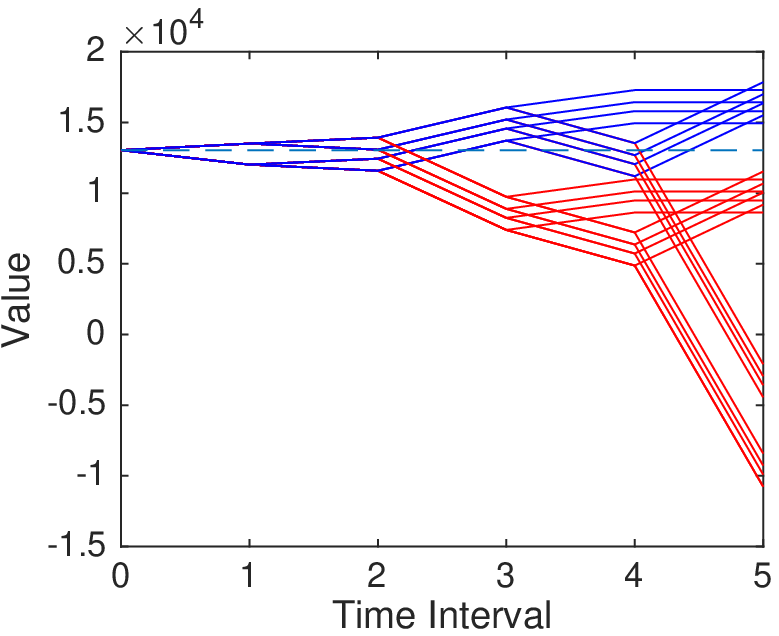}
    \end{minipage}
}
\subfigure[Full / High-low]{
    \begin{minipage}[b]{0.23\textwidth}
    \includegraphics[width=1.5in]{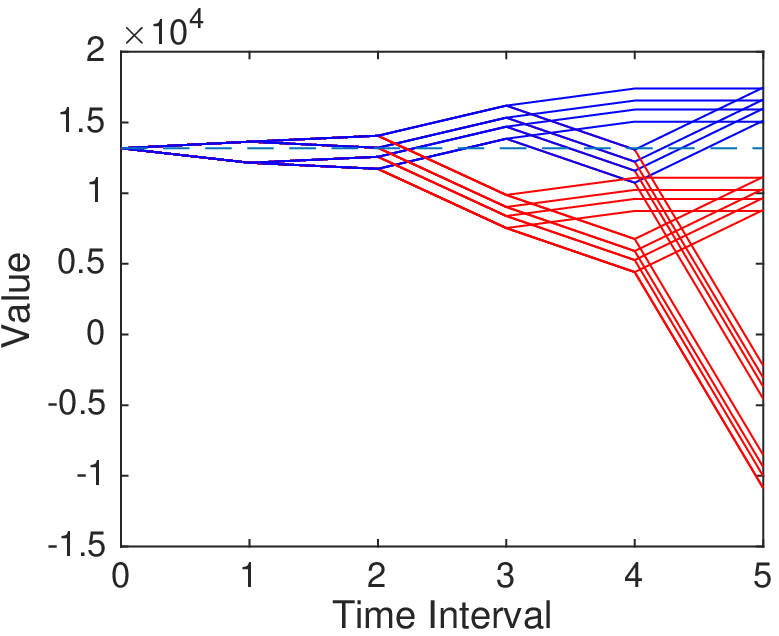}
    \end{minipage}
}

\subfigure[Small / High]{
    \begin{minipage}[b]{0.23\textwidth}
    \includegraphics[width=1.5in]{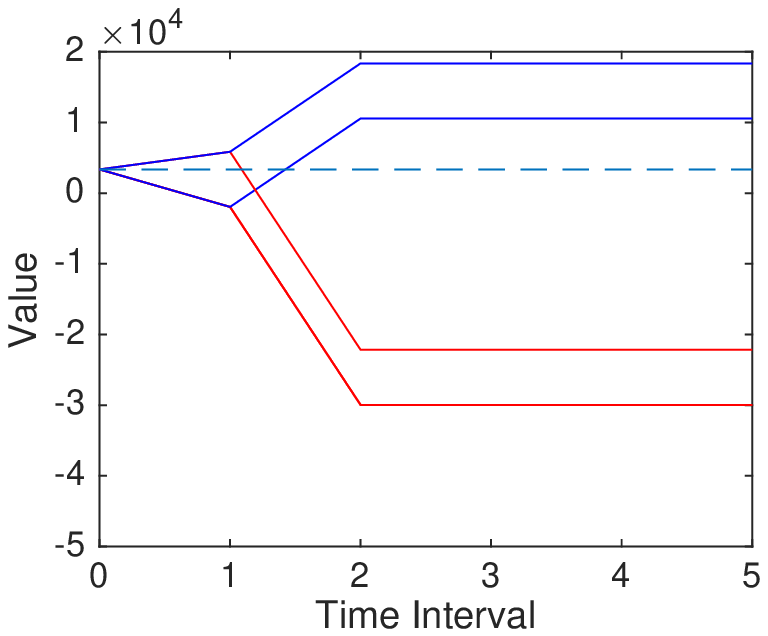}
    \end{minipage}
}
\subfigure[Medium / High]{
    \begin{minipage}[b]{0.23\textwidth}
    \includegraphics[width=1.5in]{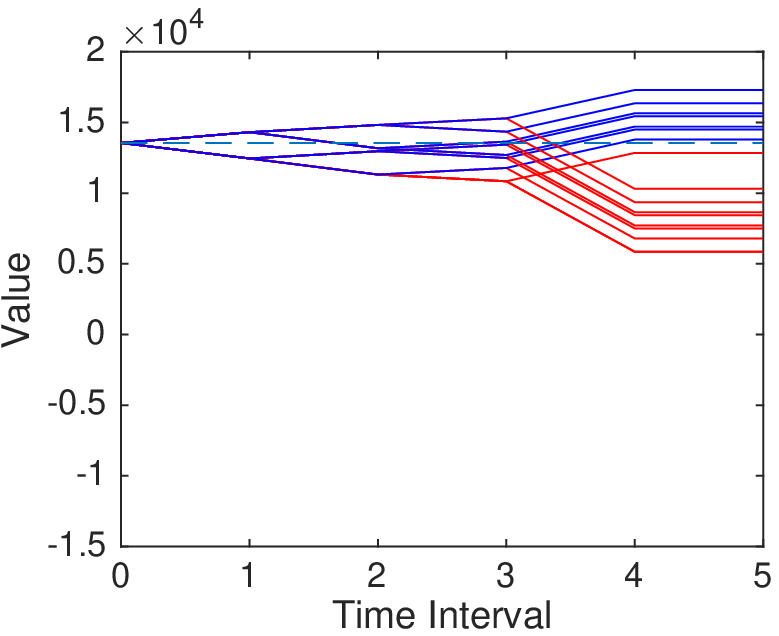}
    \end{minipage}
}
\subfigure[Large / High]{
    \begin{minipage}[b]{0.23\textwidth}
    \includegraphics[width=1.5in]{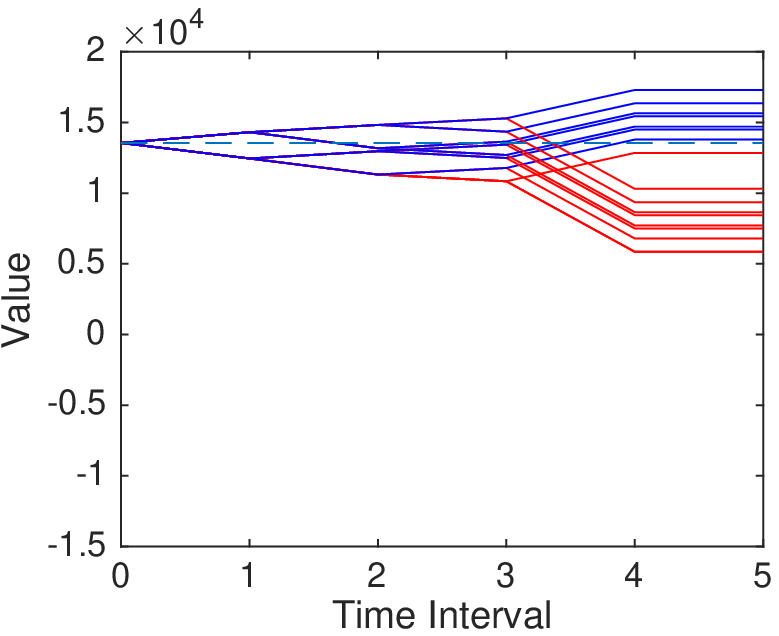}
    \end{minipage}
}
\subfigure[Full / High]{
    \begin{minipage}[b]{0.23\textwidth}
    \includegraphics[width=1.5in]{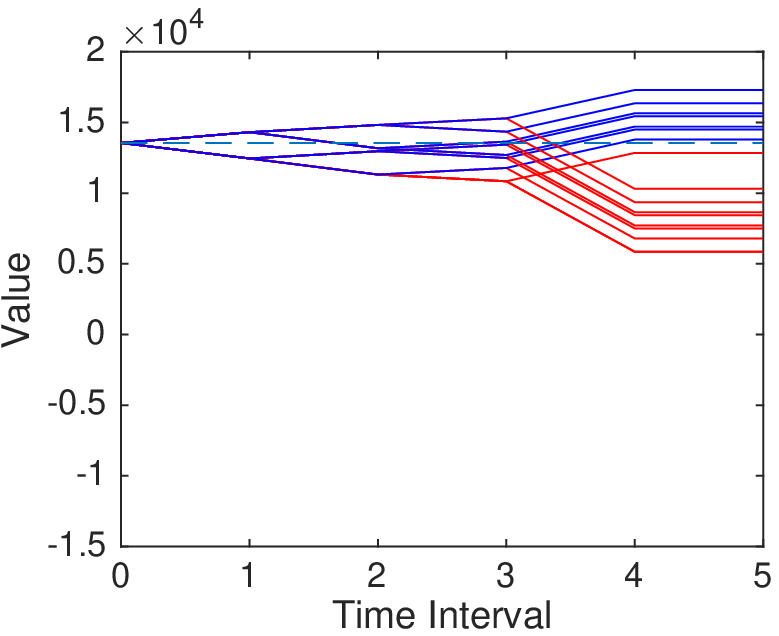}
    \end{minipage}
}

\caption{Value Plots of the 16 HVTs} 
\label{fig:ValuePlots}
\vspace{-6mm}
\end{figure*}

The evolution of the values of the different verification strategies as they progress through the time intervals is depicted in Fig.~\ref{fig:ValuePlots}. Each plot consists of a series of lines that match all possible paths for the HVT. The dashed lines show the expected values of the HVTs. The blue lines represent the paths that yield higher values than the expected value at the end $t = T$; the red lines represent those paths that yield lower values. The values yielded by each path at each time interval are the sum of the items that have happened and the expected values of the following FVTs in the following time intervals. For most cases, the values at early time intervals do not represent the value at the end, which can be observed by tracing the changes of lines.

From the 16 HVTs and their value plots, it can be found the network size has a fundamental but limited impact on the generation of strategies. When the network is small, there are at most two verification activities in the strategy because the small network contains only five activity choices. When the network become a medium one,  both the depth of trees and the number of paths increases. However, comparing the medium network with large/full networks, there are no much changes of tree shapes even though the specific selection of some nodes is different. A possible explanation is that most of the system parameters and verification activities in large/full networks not included in the medium network are farther from the target node than those of the medium network. This means that the additional nodes have limited influence on the confidence level of $P(\theta_3)$. This can also be certificated by comparing the expected values of value plots. That is, there is a sharp increase of expected values from the small network to the medium network. The expected values do not increase so much from the medium network to larger ones. This contradicts the intuition that more activities would bring more choices and potential opportunities for better results. We assume, however, that the reason for this result is that including more possible verification activities enlarges the verification tradespace and, hence, requires more calculations to find the solution.

In addition, the rework rules also significantly influence the selection of VAs. When the decision threshold is set at a high level, triggering rework activities becomes much easier. Thus the number of different paths can be reduced for a certain network size. This can be clearly seen in the last two columns of HVTs in Fig.~\ref{fig:HVTs}. From the value plots, when the rework rule is set at a high level and the network is larger than the small one, the range of path values at $t=5$ also decreases to $[5000, 40000]$. This is explained by the effect of frequent rework that can prevent more serious errors in the late time intervals. It is noticeable that frequent rework does not necessarily result in lower expected values because of the optimization process. That is, verification activities that are associated with low rework costs become much more preferred when rework is inevitable. In contrast, when the lower threshold is on the low end, those VAs that have a high impact on confidence can be tested with a lower risk of large rework costs.

\begin{figure*}
\vspace{-1mm}
\centering
\subfigure[Low]{
    \begin{minipage}[b]{0.23\textwidth}
    \includegraphics[width=1.5in]{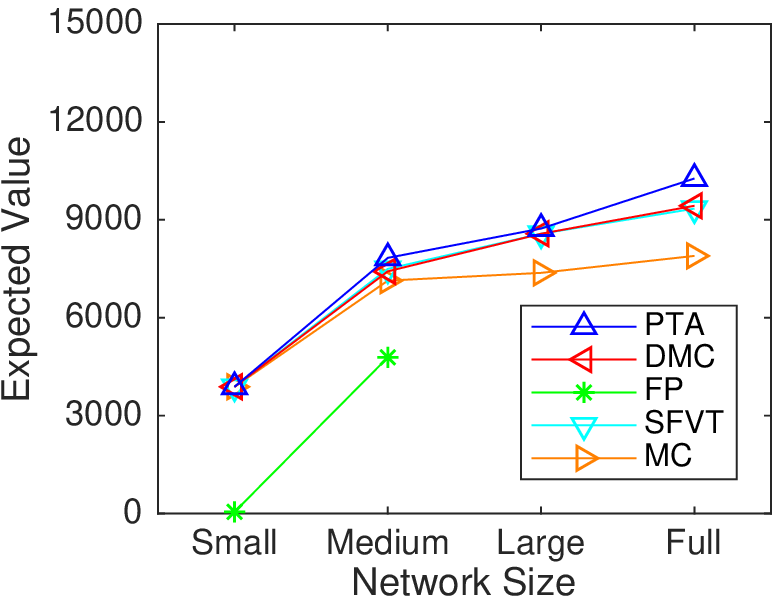}
    \end{minipage}
}
\subfigure[Low-high]{
    \begin{minipage}[b]{0.23\textwidth}
    \includegraphics[width=1.5in]{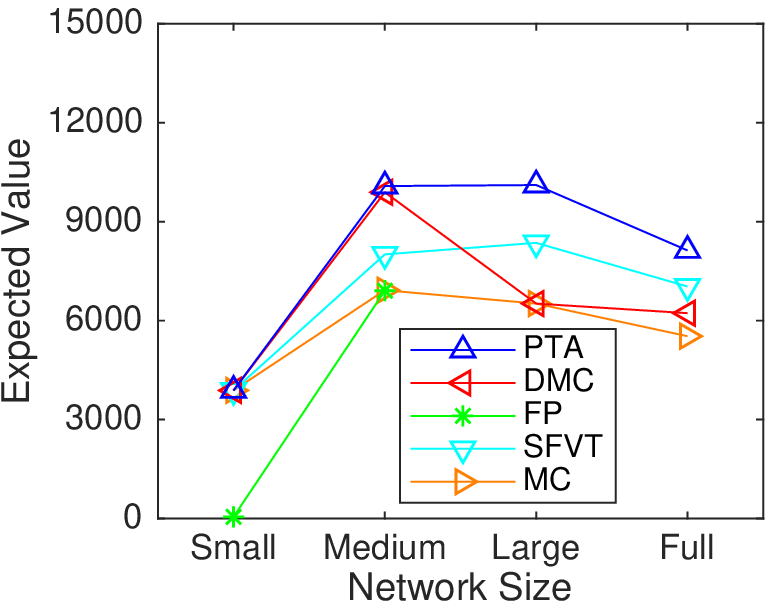}
    \end{minipage}
}
\subfigure[High-low]{
    \begin{minipage}[b]{0.23\textwidth}
    \includegraphics[width=1.5in]{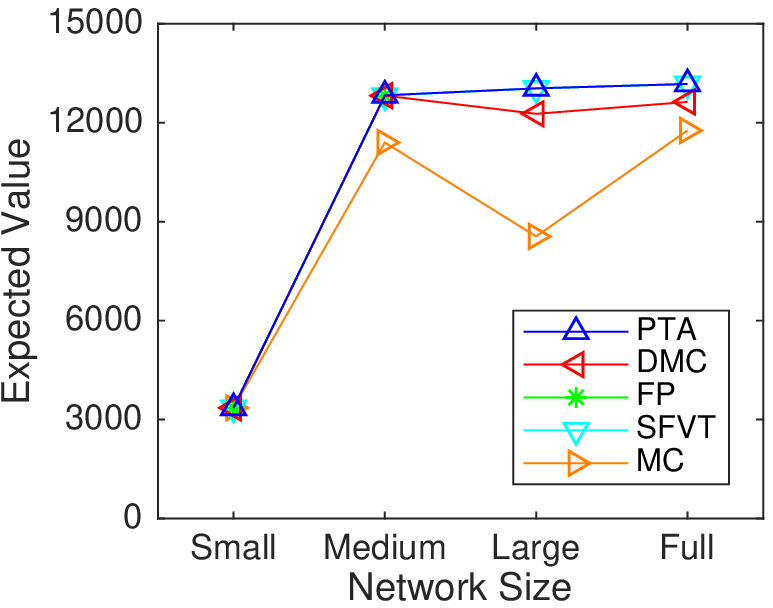}
    \end{minipage}
}
\subfigure[High]{
    \begin{minipage}[b]{0.23\textwidth}
    \includegraphics[width=1.5in]{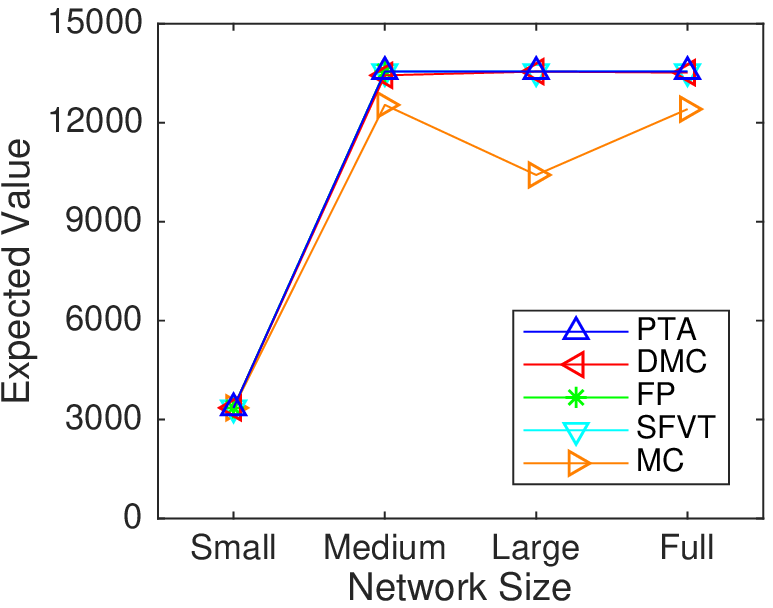}
    \end{minipage}
}

\subfigure[Low]{
    \begin{minipage}[b]{0.23\textwidth}
    \includegraphics[width=1.5in]{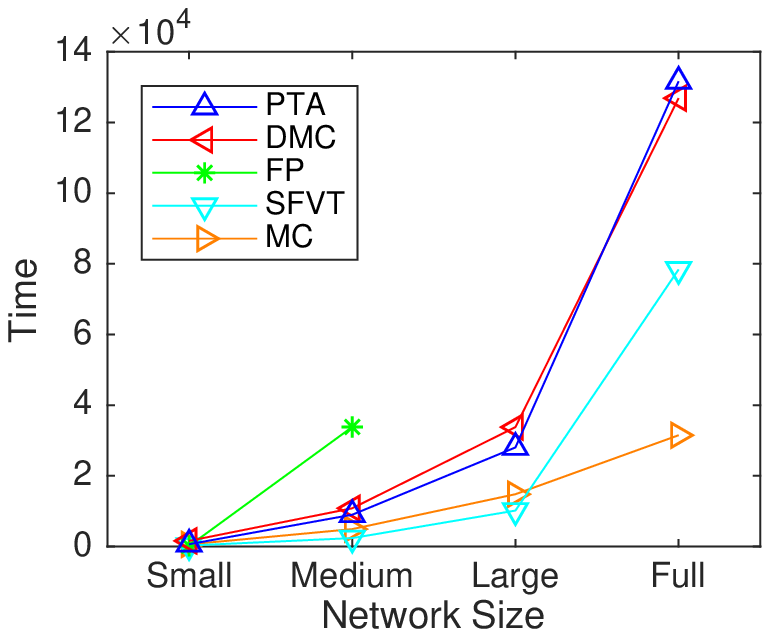}
    \end{minipage}
}
\subfigure[Low-high]{
    \begin{minipage}[b]{0.23\textwidth}
    \includegraphics[width=1.5in]{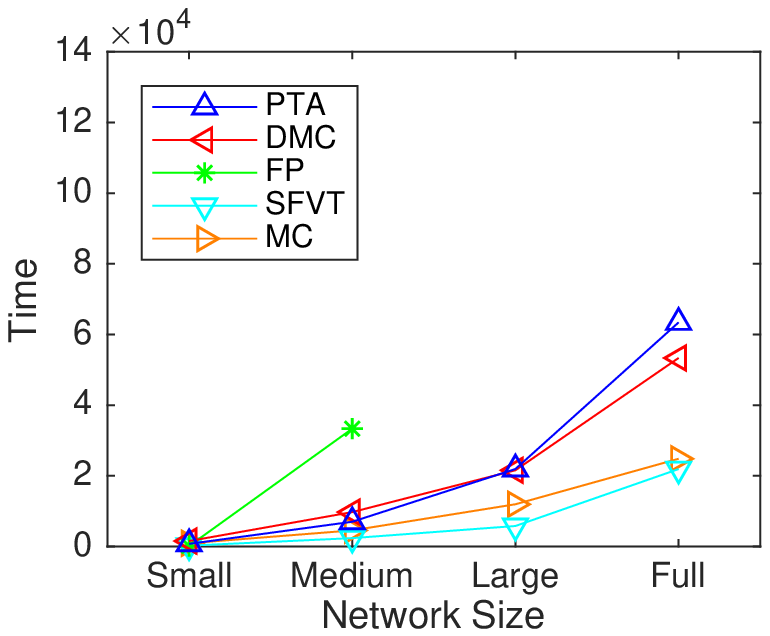}
    \end{minipage}
}
\subfigure[High-low]{
    \begin{minipage}[b]{0.23\textwidth}
    \includegraphics[width=1.5in]{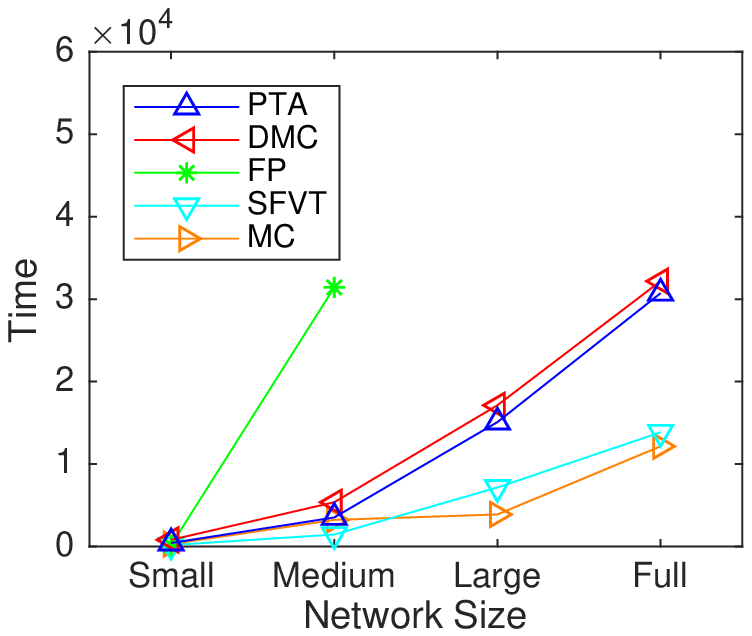}
    \end{minipage}
}
\subfigure[High]{
    \begin{minipage}[b]{0.23\textwidth}
    \includegraphics[width=1.5in]{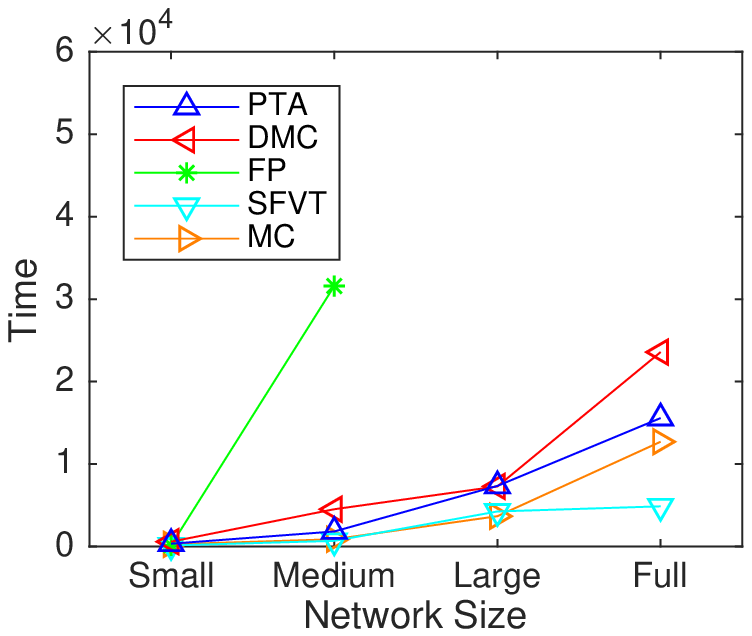}
    \end{minipage}
}
\caption{Comparison of Expected Values (a - d) and Run Time (e - h) between the Proposed Method (i.e., PTA) and Benchmark Methods (i.e., FP, MC, DMC, and SFVT)}
\label{fig:ComparisonPlot}
\vspace{-5mm}
\end{figure*}

The expected value comparison between the proposed method and benchmark methods are listed in Table~\ref{tab:methodcomparision} and Fig.~\ref{fig:ComparisonPlot} (a)-(d). Notably, as there are many more system parameters and VAs in the large and full networks, the brute force-based FP method cannot be applied in these two networks. Results show that the expected values of the verification strategies yielded by PT-based methods (i.e., PTA and SFVT) is always better than those of the verification strategies yielded by the brute force-based and Monte Carlo-based methods (i.e., FP, DMC, MC). It can be found dynamic design could also enhance the performance to some extent by comparing PTA with SFVT and comparing DMC and MC. The proposed PTA always yields the highest expected value among the different methods tested. Superiority, however, is marginal for the small network or the `High-low' and `High' rework rules. A possible explanation for this performance is that the network size or rework mechanism causes a shrinkage of the verification tradespace. For example, when the `High' rework rule is applied, the rework is almost always triggered. As a consequence, the resulting tree tradespace becomes a path tradespace that has no more than $5$ fixed activities. While there is up to $10*9^2*8^4*7^8*6^{16}=5.40 \times 10^{25}$ tree solutions with a 31-node tree structure (i.e., $1+2+4+8+16=31$ dimensions), the path tradespace has no more than $10*9*8*7*6=30,240$ solutions. Obviously, exploration is much easier for the latter example. 

Run time results of the methods in comparison are listed in Table~\ref{tab:timecomparision} and Fig.~\ref{fig:ComparisonPlot} (e)-(h). When the network scales from small to full, the order of time magnitude of all methods increases significantly. So it can be concluded that network size has a direct impact on run time. From the aspects of methods, the FP method is more efficient than others when the network is small. But as the network size increases to a medium one, the FP method is the most time-consuming. This is explained by the inherent brute-force method that all possible strategies must be examined. The rest of the methods maintain the same order of time magnitude for each case. In particular, the PT-based methods (i.e, PTA, SFVT) is slightly better than MC-based methods (i.e., MC, DMC), especially when the network size is small or medium.


\begin{table}[]
\centering
\caption{Expected Value Comparison between the Proposed Method (i.e., PTA) and Benchmark Methods (i.e., FP, MC, DMC, and SFVT) (unit: \$1,000)}
\label{tab:methodcomparision}
\scriptsize 
\begin{tabular}{|c|c|cc|ccc|}
\hline
\makecell*[c]{Rework\\Rule} & \makecell*[c]{Network\\Size} & \makecell*[c]{PTA} & \makecell*[c]{DMC} & \makecell*[c]{FP} & \makecell*[c]{SFVT} & \makecell*[c]{MC} \\ \hline
\hline
\multirow{4}{*}{Low} & Small & 3,884 & 3,884 & 50 & 3,884 & 3,884 \\
 & Medium & \textbf{7,835} & 7,422 & 4,786 & \textbf{7,497} & 7,137 \\ 
 & Large & \textbf{8,733} & 8,574 & - & \textbf{8,574} & 7,375 \\
 & Full & \textbf{10,267} & 9,431 & - & \textbf{9,344} & 7,895 \\ \hline
 \hline
\multirow{4}{*}{Low-high} & Small & 3,884 & 3,884 & 50 & 3,884 & 3,884 \\
 & Medium & \textbf{10,080} & 9,892 & 6,907 & \textbf{8,009} & 6,922 \\ 
 & Large & \textbf{10,111} & 6,517 & - & \textbf{8,358} & 6,517 \\
 & Full & \textbf{8,128} & 6,226 & - & \textbf{7,035} & 5,528 \\ \hline
 \hline
\multirow{4}{*}{High-low} & Small & 3,356 & 3,356 & 3,356 & 3,356 & 3,356 \\
 & Medium & 12,832 & 12,832 & 12,821 & 12,821 & 11,401 \\ 
 & Large & \textbf{13,042} & 12,270 & - & \textbf{13,042} & 8,554 \\
 & Full & \textbf{13,175} & 12,631 & - & \textbf{13,175} & 11,765 \\ \hline
\hline
\multirow{4}{*}{High} & Small & 3,356 & 3,356 & 3,356 & 3,356 & 3,356 \\
 & Medium & \textbf{13,551} & 13,437 & 13,551 & 13,551 & 12,541 \\ 
 & Large & 13,551 & 13,551 & - & \textbf{13,551} & 10,417 \\
 & Full & \textbf{13,551} & 13,519 & - & \textbf{13,551} & 12,412 \\ \hline
\end{tabular}%
\vspace{-2mm}
\end{table}

\begin{table}[]
\centering
\caption{Run Time Comparison between the Proposed Method (i.e., PTA) and Benchmark Methods (i.e., FP, MC, DMC, and SFVT) (unit: second)}
\label{tab:timecomparision}
\scriptsize
\begin{tabular}{|c|c|cc|ccc|}
\hline
\makecell*[c]{Network\\Size} & \makecell*[c]{Rework\\Rule} & \makecell*[c]{PTA} & \makecell*[c]{DMC} & \makecell*[c]{FP} & \makecell*[c]{SFVT} & \makecell*[c]{MC} \\ \hline
\hline
\multirow{4}{*}{Low} & Small & \textbf{649} & 1,595 & \textbf{163} & 277 & 607 \\
 & Medium & \textbf{8,999} & 10,826 & 33,827 & \textbf{2,383} & 4,927 \\ 
 & Large & \textbf{28,036} & 33,784 & - & \textbf{10,173} & 14,802 \\
 & Full & 131,625 & \textbf{126,870} & - & 78,393 & \textbf{31,489} \\ \hline
 \hline
\multirow{4}{*}{Low-high} & Small & \textbf{752} & 1,551 & \textbf{148} & 241 & 923 \\
 & Medium & \textbf{7,009} & 9,741 & 33,347 & \textbf{2,375} & 4,563 \\ 
 & Large & 21,936 & \textbf{21,584} & - & \textbf{5,807} & 11,958 \\
 & Full & 63,413 & \textbf{53,353} & - & \textbf{21,917} & 24,829 \\ \hline
 \hline
\multirow{4}{*}{High-low} & Small & \textbf{428} & 809 & \textbf{153} & 172 & 305 \\
 & Medium & \textbf{3,523} & 5,369 & 31,422 & \textbf{1,455} & 3,228 \\ 
 & Large & \textbf{15,090} & 17,130 & - & 7,160 & \textbf{3,888} \\
 & Full & \textbf{30,723} & 32,181 & - & 13,852 & \textbf{12,155} \\ \hline
\hline
\multirow{4}{*}{High} & Small & \textbf{340} & 586 & 155 & \textbf{130} & 267 \\
 & Medium & \textbf{1,824} & 4,522 & 31,592 & \textbf{686} & 883 \\ 
 & Large & 7,351 & \textbf{7,291} & - & 4,251 & \textbf{3,677} \\
 & Full & \textbf{15,576} & 23,576 & - & \textbf{4,874} & 12,706 \\ \hline
\end{tabular}%
\vspace{-3mm}
\end{table}


\subsection{Discussion}
\label{subsec:Discussion}
We would like to make several remarks for the proposed methodology and the obtained experimental results.

First, from \Cref{tab:methodcomparision,tab:timecomparision} and Fig.~\ref{fig:ComparisonPlot}, it is found that the proposed PTA outperforms the benchmark methods considering both expected value and run time.
This advantage is mainly attributed to the dynamic design and the PT feature of continually optimizing the sub-optimal solutions. To be more specific about the PT feature, as the PT-based methods always generates similar samples to previous ones, it can consistently explore around the certain sample spaces before jumping out to another space, especially in the low temperature replicas~\citep{Earl2005}. In contrast, the MC method always generates a completely new sample at each iteration. Even though the MC method can jump out of the local optimum quickly, it lacks the ability to exploit the promising space. The experimental results also show that the benchmark methods have their own merits. The FP method is more efficient than the other methods when the network is small. The HVTs of the FP also have the highest expected value when the `High' rework rule is applied. Compared with the PTA, the SFVT is an economic choice to solve the strategy design problem, especially when the dimensions of a tradespace is small. The reason is its run time is always less than one half of the PTA and the expected values of its HVTs are close to that of the PTA. 


\begin{figure}[htbp]
\vspace{-3mm}
\centerline{\includegraphics[width=8cm]{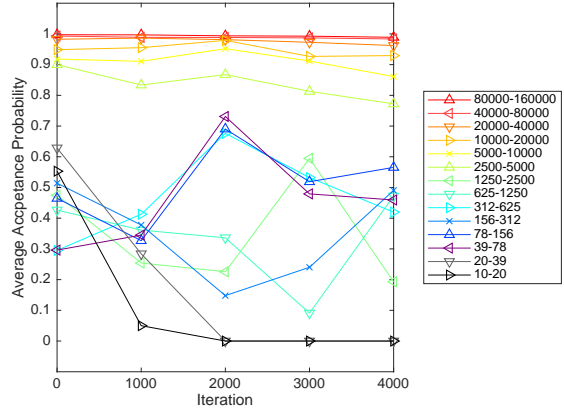}}
\caption{Average Acceptance Probabilities of All Neighboring Replicas (e.g., `10-20' represents the swap of the two replicas $\{\Omega(\Psi_1=10),\Omega(\Psi_2=20)\}$.)}
\label{fig:AcceptRate}
\vspace{-3mm}
\end{figure}

\begin{figure}[htbp]
\vspace{-2mm}
\centerline{\includegraphics[width=6cm]{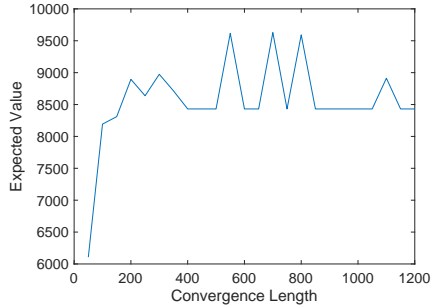}}
\caption{Sensitivity Test for the Convergence Length}
\label{fig:ConvergenceLength}
\vspace{-5mm}
\end{figure}

Second, as presented in Section~\ref{subsec:paralleltempering}, 
there are several important parameters for designing the PT, including temperatures and convergence length. The parameters need further evaluation in terms of the computing performance after the experiment. First, the spacing between temperatures can be tuned by evaluating the acceptance probability. We calculate the average acceptance probability every 20 swaps (i.e., 1000 iterations) of all neighboring replicas for Case (d), which has the largest tradespace. For example, the swap between the two replicas $\{\Omega(\Psi_1=10),\Omega(\Psi_2=20)\}$ is denoted as `10-20', as shown in Fig.~\ref{fig:AcceptRate}. The two lowest pairs `10-20', `20-39' become $0$ after the first 2000 iteration. This is because the best configuration is swapped to the lowest temperature replica at iteration 1600 and stay there afterwards. Most of other pairs are larger than the recommended value of $0.2$~\citep{rathore2005optimal}. So the configurations can be accepted between high and low temperature replicas actively. Second, a sensitivity analysis is made for the convergence length $L$. We test the effect of various length values with a unit of 50 for Case (d). As shown in Fig.~\ref{fig:ConvergenceLength}, when the length is larger than 200 iterations, the expected value is no less than 8,430. To ensure a sufficient redundancy, its value was set as 1,000 so that there was enough time to find a better solution before the iteration process converged. As the PTA is a heuristic method, it is still possible to generate an unsatisfactory solution. However, we assume that with this length, the probability of this phenomenon could be controlled within a certain range.

Third, an implicit result of verification strategies can be found through the comparison of the generated HVTs in Fig.~\ref{fig:HVTs}. Note that $A_{24}$ is always conducted at the end of verification processes when the confidence $P(\theta_3)$ is high enough. As $A_{24}$ has larger rework costs than other activities, it is riskier to conduct this type of activities if they are likely to trigger rework activities after collecting their results. Therefore, conducting low-risk activities first may yield more information about the engineered system being verified, while reducing the probability that rework happens. For example, the prior probability $P(A_{24}=F)$ is $0.396$. But if $A_{38}$ is implemented first and its result is $A_{38}=T$, as shown in case (d), $P(A_{24}=F|A_{38}=T)$ will decrease to $0.334$. As a result, rework is less likely to be triggered. From a practitioner's standpoint, this result can be interpreted as prioritizing verification activities that quickly increase confidence at low risk of rework, gradually incorporating high risk activities as the confidence on the correct operation of the system increases.

\section{Conclusion}
\label{sec:Conclusion}
In this paper, we present a parallel tempering approach to explore high-dimension verification tradespaces for engineered systems. This approach follows the need to apply SBD to the design of verification strategies. When considering dynamic verification strategies, the exploration problem of near-optimal verification activities is formulated as a tree search problem. Then, we designed the PT algorithm with the characteristics of verification processes. The experiments are designed with four networks of different sizes and four rework rules. 

The experiments show that the proposed PTA outperforms the benchmark methods in most cases. Its scalability in network size is also justified by comparing four networks. The expected values of the verification strategies yielded by the PTA are always better than those achieved when using baseline methods, especially in high dimension tradespaces. In terms of computational efficiency, the proposed PTA outperforms current approaches based on enumeration when the network is large. PTA also shows its advantage in low dimension tradespaces, and is on par with the other benchmark methods in high dimension tradespaces. We suggest that adding features or rules about the system of interest could accelerate optimization. 

It is also important to note that the proposed method has been designed with certain assumptions for simplicity. First, we assumed that Bayesian networks can fully capture the confidence relationships of engineered systems and verification strategies, which may be hard to realize in reality. Second, predefined rework and system deployment rules against confidence thresholds have been used instead of determining optimal actions. Third, the values of all parameters are assumed to be fixed for all cases. More adaptive mechanisms can be added to accelerate the PT process as a future work. Nevertheless, we suggest that these assumptions are reasonable within the context of the work presented in this paper. 

Estimating these values (e.g., cost values and rework thresholds), while important, were left outside of the scope of the paper. Yet, we offer some informative (non-prescriptive) guidance for how they may be calculated. Estimating verification setup costs is common in practice. Proprietary parametric cost models that are built using historical data could be used to create initial, rough estimations. Direct proposals from vendors and service providers, which require more effort to obtain, may be used to refine and/or increase the confidence of the estimates. Estimating rework costs and rework thresholds is less straightforward. Rough estimations of rework costs may be obtained by leveraging historical data as a function of those incurred at different milestones in the development process. To improve estimation confidence, adequate tasking, planning, and resource allocation (in terms of personnel, material, and facility/equipment) could be used to identify those tasks that would need to be repeated for each verification node, should a rework decision be made for that particular node. Because rework thresholds are used as pre-defined rework decisions based on achieved confidence level, we suggest to establish them using utility theory. Specifically, the confidence level could be set by finding the expected consequence of carrying on a system error (as a function of the confidence), adjusted with the risk profile specific for the project, that is equivalent to the expected cost of rework. While this approach does not guarantee optimality (for that, rework thresholds should be substituted by dedicated rework decisions), we believe that it offers a sufficiently good approximation while making it feasible for adoption in practice.

\section*{Acknowledgment}

This material is based upon work supported by the Acquisition Research Program under HQ00341810002. The views expressed in written materials or publications, and/or made by speakers, moderators, and presenters, do not necessarily reflect the official policies of the Department of Defense nor does mention of trade names, commercial practices, or organizations imply endorsement by the U.S. Government.

The authors thank Advanced Research Computing at Virginia Tech for providing the parallel computing services.

\bibliographystyle{unsrtnat}
\bibliography{PT}  






\end{document}